\documentclass[twocolumn]{aastex62}
\usepackage{graphicx}			
\usepackage{amsmath,amssymb}
\usepackage[caption=false]{subfig}
\usepackage{soul}
\usepackage{booktabs}
\usepackage{scrextend}
\usepackage{xcolor}

\newcommand{\msun}{M_\odot}
\newcommand{\mzams}{M_{\rm ZAMS}}
\newcommand{\minmass}{M_\textrm{min}}
\newcommand{\maxmass}{M_\textrm{max}}
\newcommand{\tmin}{t_\textrm{min}}
\newcommand{\tmax}{t_\textrm{max}}
\newcommand{\minmassvalue}{7.33^{+0.02}_{-0.16}}
\newcommand{\maxmassvalue}{59}
\newcommand{\alphavalue}{-2.96^{+0.45}_{-0.25}}
\newcommand{\minage}{4.2}
\newcommand{\maxage}{50.3^{+2.5}_{-0.5}}
\newcommand{\betavalue}{0.38^{+0.18}_{-0.32}}
\newcommand{\nsnr}{94}
\newcommand{\nsnrmto}{62}
\newcommand{\nsnrmtt}{32}

\newcommand{\RomanNumeralCaps}[1]
{\MakeUppercase{\romannumeral #1}}

\received{2018 February 15}
\revised{2018 May 18}
\accepted{2018 May 18}
\published{2018 July 9}

\shorttitle{Progenitor Mass Distribution for CC SNRs}
\shortauthors{D\'iaz-Rodr\'iguez et al.}
\begin{document}
	
\title{Progenitor Mass Distribution for Core-collapse Supernova Remnants in M31 and M33}

\author[0000-0002-4652-5983]{Mariangelly D\'iaz-Rodr\'iguez}
\altaffiliation{md14u@my.fsu.edu}
\affil{Department of Physics, Florida State University, Tallahassee, FL 32304, USA}

\author{Jeremiah W. Murphy}
\altaffiliation{jwmurphy@fsu.edu}
\affil{Department of Physics, Florida State University, Tallahassee, FL 32304, USA}

\author{David A. Rubin}
\affiliation{Space Telescope Science Institute, Baltimore 21218, USA}

\author{Andrew E. Dolphin}
\affiliation{Raytheon, Tucson, AZ 85734, USA}

\author{Benjamin F. Williams}
\affiliation{Department of Astronomy, Box 351580, University of Washington, Seattle, WA  98195, USA}

\author{Julianne J. Dalcanton}
\affiliation{Department of Astronomy, Box 351580, University of Washington, Seattle, WA  98195, USA}

\begin{abstract}
	Using the star formation histories (SFHs) near $\nsnr$ supernova
	remnants (SNRs), we infer the progenitor mass distribution
	for core-collapse supernovae. We use Bayesian inference and model each SFH
	with multiple bursts of star formation (SF), one of which is assumed to be
	associated with the SNR. Assuming single-star evolution, the
	minimum mass of CCSNe is 
	$\minmassvalue$~$\msun$, the slope of the progenitor mass
	distribution is $\alpha = \alphavalue$, and the maximum mass is
	greater than $\maxmass > \maxmassvalue$~$\msun$ with a 68~\% confidence. While these results
	are consistent with previous inferences, they also provide tighter
	constraints. The progenitor distribution is somewhat steeper than
	a Salpeter initial mass function ($\alpha$~=~-2.35).  This
	suggests that either SNR catalogs are biased against the youngest SF regions,
		or the most massive stars do not explode as easily as lower mass stars.
	If SNR catalogs are biased, it will most likely 
	affect the slope but not the minimum mass. The uncertainties are dominated by three primary sources of uncertainty, the SFH resolution, the number of SF bursts, and the uncertainty on SF rate in each age bin. We address the first two of these uncertainties, with an emphasis on multiple bursts.  The third will be addressed in future work.
\end{abstract}



\section{Introduction}
\label{sec:introduction}

One fundamental prediction of stellar evolution theory is that the zero-age-main-sequence mass ($\mzams$) of a star determines its fate \citep{nomoto1987,woosley2002,heger2003}. In particular, theory predicts that single stars above $\sim$8 $\msun$ eventually collapse \citep{woosley2002}, but it is not clear if every core collapse leads to explosion. Recent investigations suggest that lower mass stars may explode more easily than higher mass stars \citep{radice2017}, with the latter being more likely to collapse directly into a black hole \citep{ugliano2012,bruenn2013,burrows2016}. However, the ease of
explosion may not be monotonic with mass
\citep{woosley2015,sukhbold2016}. For example, \citet{sukhbold2016} studied the explodability of progenitor stars from 9 to 120~$\msun$. They found that there is no clear threshold of unsuccessful SN explosion, but that stars less massive than $\sim$15~$\msun$ tend to explode. On the other hand, the region above 15~$\msun$ shows both successful explosions and failed SNe.  While the more massive stars are more likely to fail, there is not a monotonic trend.  Instead, there appear to be islands of SN production. These ``islands'' of SN production complicate the mapping between progenitor mass and explosion outcome. Moreover, the final core structure of progenitors may be
chaotic, further breaking the simple mapping of progenitor mass to outcome mapping \citep{sukhbold2017}. To constrain these basic predictions of stellar evolution, it is important to observationally constrain the progenitor mass distribution for SNe. 

Broadly, there are two methods for constraining the progenitor masses of CCSNe. One is to analyze images of the progenitor star taken before the SN \citep{white1987,smartt2002em,smartt2002ap,Smartt2004,smarttetal2009,vandyk2003du,vandyk2003gd,vandyk2011,vandyk2012a,vandyk2012rsg,li2005,li2006,li2007,maundsn2005cs,maund2011,maund2014bk,maund2014latetime,hendry2006,galyam2007,galyam2009,smartt2009,smith2011,fraser2012,fraser2014}. This technique has the advantage that it directly probes the progenitor star, allowing the identification of the type of star that exploded, and to infer its $\mzams$, one compares the color and magnitude of the precursor to stellar evolution tracks.

\begin{figure*}
	\includegraphics[width=1\textwidth]{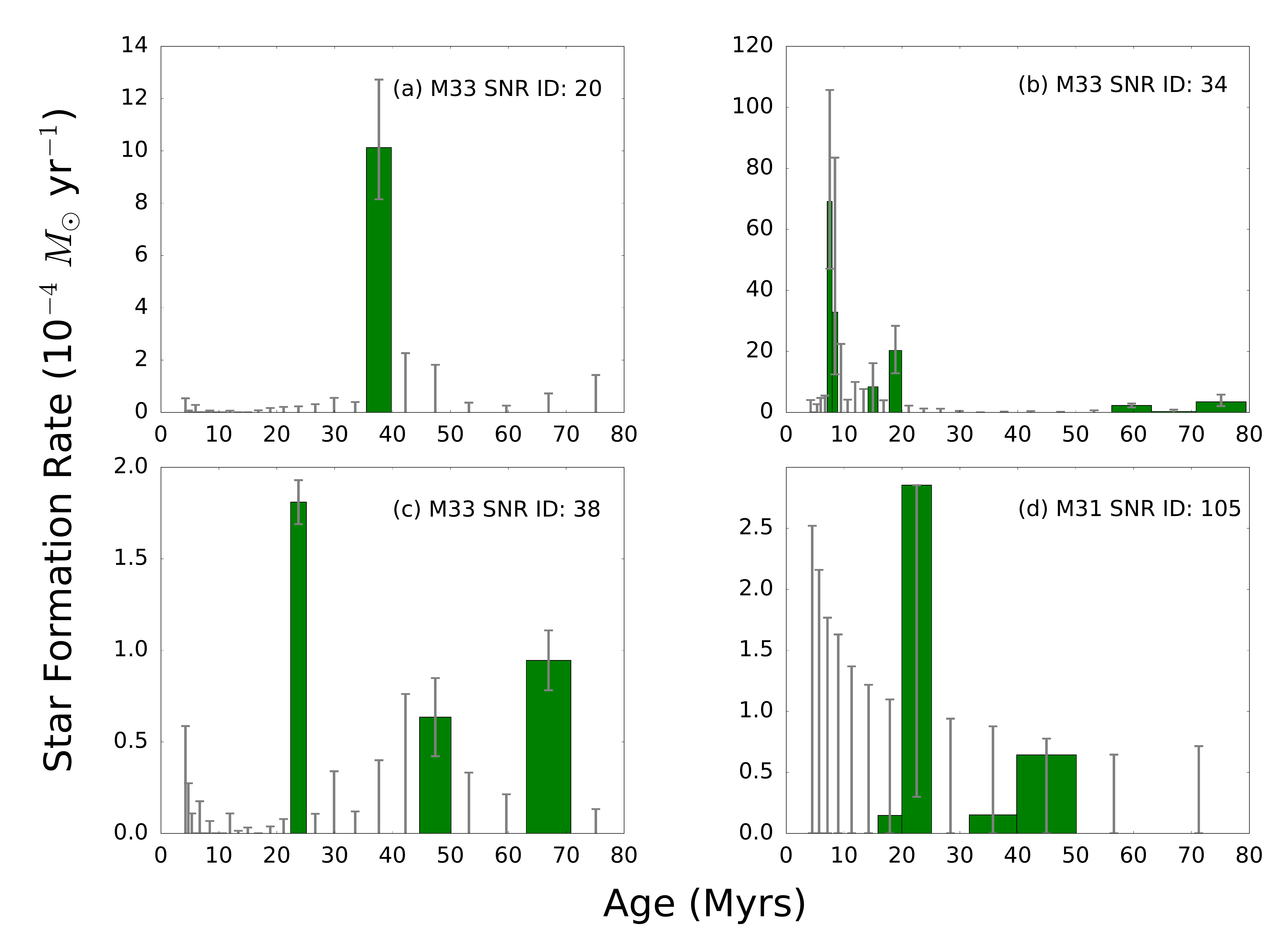}
	\caption{The star formation history (SFH) within 50 pc of four SNRs
		in M31 and M33: Sometimes, (a) the SFH identifies one clear age (one burst of SF). However, often (b)-(d)
		there is more than one burst of SF in the last 80 Myr.  Only
		one burst in each SNR is associated with the progenitor that exploded. The others are
		random unassociated bursts of SF. For some, the
		SFH has many young bright main-sequence stars and the SFH is
		quite certain (a), and for others there are very few bright main-sequence stars within 50 pc, and the SFH is quite uncertain
		(d).} 
	\label{fig:data}
\end{figure*}

 Moreover, even when precursor images exist inferring the luminosity and mass of the progenitor requires interpreting a star magnitude and color during the last, most uncertain stages of stellar evolution, for which the star may not be in hydrostatic equilibrium \citep{quataertandshiode2012,fuller2017}. In addition, these stars' dusty winds may be obscuring their luminosity and consequently lowering mass estimates \citep{vandyk2012rsg,beasor2016,beasor2017}. 
 
While the direct-imaging method constrains the type of star that
exploded, there are potential limitations to the technique. For one,
serendipitous precursor images are rare. As a result, only about 30 SNe
have directly imaged progenitors and 38 upper limits \citep{vandyk2017}.
One of the largest analyses of precursor images \citep{smartt2015}
finds that all SN II-P direct detections were red super giants (RSGs),
as expected. They also infer that the minimum $\mzams$ for explosion
is $7^{+4}_{-1}\msun$, observed for SN2003gd \citep{smarttetal2009,Smartt2004}. Surprisingly, even though RSGs are observed with
masses up to 25-30~$\msun$, there are no SN II-P progenitors more massive than $\sim$ 17~$\msun$ \citep{smartt2015}. Recently, however, \citet{daviesbeasor2017} applied different bolometric corrections that are more appropriate for late-stage RSGs and found a higher upper mass limit of $<$ 27 $\msun$ (95\% confidence). Therefore, another technique with different systematics and limitations is needed to increase the number of progenitors when pre-images are not available and to validate the direct-imaging method.


The second alternative technique is to age-date the surrounding stellar populations in
the vicinity of the SN explosion, and from this age, infer a progenitor mass \citep{walborn1993, barth1996,vandyk1999,panagia2000, maizapellaniz2004, wang2005,crockett2008, gogarten2009,vinko2009,murphy2011,murphy2018,williams2014,williams2018,maund2017,maund2018}. The age-dating technique mostly depends on well-understood properties of main-sequence and early post-main-sequence phases, and thus is relatively insensitive to details of late-stage stellar evolution.

Because this age-dating technique does not require rare precursor imaging, age-dating expands the number of progenitor estimates to many more CCSNe \citep{williams2014,williams2018,maund2017,maund2018} and to hundreds of supernova remnants (SNRs; \citealt{jennings2012,jennings2014}). For example, \citet{williams2018} used this technique to
age-date 25 historic SNe.
\citet{maund2017} used a similar age-dating
	technique to resolve stellar populations around 12 Type II-P SNe
	with identified progenitors, and \citet{maund2018} inferred the
	ages around the sites of 23 stripped-envelope SNe. \citet{jennings2014} age-dated 115 SNRs, demonstrating a way to increase the number of progenitor masses by at least a factor of 10. 

\citet{jennings2012,jennings2014} preliminarily constrained the minimum mass ($\minmass$), maximum mass ($\maxmass$), and power-law slope ($\alpha$) of the progenitor mass distribution, assuming SNRs were unbiased tracers of recent SNe. \citet{jennings2012,jennings2014} were not able to infer all three parameters simultaneously, and instead employed several models to constrain the distribution using KS statistics. In a smaller initial sample, \citet{jennings2012} found a $\minmass$ for CCSNe between 7.0 and 7.8~$\msun$. Fixing the power-law slope to 2.35 (Salpeter IMF), \citet{jennings2014} found a $\maxmass$ of $35^{+5}_{-4}$ for an expanded sample. If instead, they assumed no $\maxmass$, they found a steeper power-law slope of $4.2^{+0.3}_{-0.3}$. In either model, they found that either the most massive stars are not exploding at the same frequency as lower masses, or there is a bias against SNRs in the youngest regions.

Initial results from age-dating have been promising, but these preliminary analyses could be improved in two ways. First, \citet{jennings2014} adopted one median age and uncertainty for each SNR, which is only appropriate if there is one well-defined peak for the SFH. In contrast, the data is often consistent with there being more than one burst of SF (see~Figure~\ref{fig:data}). Second, \citet{jennings2014} did not infer the $\minmass$, $\maxmass$, and the power-law slope simultaneously. To appropriately infer these parameters, one needs to fit for all of them at the same time.


In this paper, we begin building a complete statistical inference framework that handles these previous limitations. Here, we use a Bayesian inference framework, to infer the parameters of the progenitor age distribution simultaneously, taking multiple bursts of SF into account. Instead of focusing on masses directly, we first infer the minimum age ($\tmin$), maximum age ($\tmax$), and slope of the age distribution ($\beta$). We then use the results of stellar evolution models \citep{marigo2017} to infer the progenitor mass distribution associated with this age distribution.
  
An outline of the paper is as follows. In Section~\ref{section:methods}, we present a Bayesian inference technique to
infer the CCSN progenitor age distribution. This section also
describes the assumptions and technique to transform this age distribution into a mass distribution. Section~\ref{section:results} presents the results. In Section~\ref{section:discussion}, we discuss our results in the context of other progenitor analyses, theory, and major potential biases. We summarize our results in Section~\ref{section:conclusion}, and discuss future directions.

\hfill
\section{Methods}
\label{section:methods}
This section describes the methods for inferring the progenitor
age and mass distribution for SNRs in M31 and M33.  The primary inference is the age
distribution rather than the mass distribution for several reasons.
First, the fundamental result for each SNR is the age of the local
stellar population.  Second, to infer the progenitor mass
distribution, one makes assumptions about the mapping from age to
mass.  The most basic mapping assumes single-star evolution.  However,
binary evolution can significantly affect this mapping.  Therefore, we
first infer the progenitor age distribution assuming single-stellar evolution to allow future
investigations using binary evolution.  Since it is a standard
assumption, we then convert the progenitor age distribution into a mass
distribution. 

The methods are presented as follows.
Section~\ref{section:SNRcatalogs} briefly describes the selection
criteria for the M31 and M33 SNR catalogs; we include brief
discussions on how these selections may impose biases in the SNR
catalog.  Section~\ref{section:SFH} briefly describes the method for
inferring the SFHs from Hubble Space Telescope (HST) photometry for each SNR.  In section~\ref{subsection:agedistribution}, we describe the method for converting each SFH into an
age probability density function (PDF) for each SNR. Then, in
section~\ref{sec:hierarchical}, we use hierarchical  Bayesian inference
to infer the progenitor age distribution for all SNRs, and convert
this age distribution into an initial mass distribution for all SNRs. In section~\ref{sec:fakedata}, we simulate data to test the hierarchical model and to identify any biases.

\subsection{SNR Catalogs}
\label{section:SNRcatalogs}

We analyze the age distribution of $\nsnr$ SNRs for M31 \citep{leeandlee2014} and M33  \citep{long2010} that also have high quality overlapping HST imaging. $\nsnrmto$ of the SNRs in our analysis are in M31 and the rest are in M33.

The M31 SNR candidates were selected based on their their [S \RomanNumeralCaps{2}]:H$\alpha$, morphology, and the absence of blue stars \citep{leeandlee2014}. Their primary motivation in omitting objects with blue stars was to remove H \RomanNumeralCaps{2} regions from the catalog. However, this decision may bias against including SNRs associated with the youngest stellar populations.  In contrast, the M33 SNR candidates were selected only based on their elevated [S~\RomanNumeralCaps{2}]:H$\alpha$, regardless of size or morphology \citep{long2010}. One of the disadvantages with the M31 data is that there is very little follow-up spectroscopy, in contrast with M33. However, \citet{leeandlee2014} had the benefit of earlier surveys, and they covered the entire disk of M31. It is possible that they may have included some very faint objects that may not be SNRs. Nevertheless, \citet{leeandlee2014}'s catalog is the best extragalactic SNR survey available at the moment in comparison with other M31 catalogs. For example, the \citet{magnier1995} M31 catalog did not include the [S \RomanNumeralCaps{2}]:H$\alpha$ criteria. This criteria is very important for identifying SNRs since elevated [S \RomanNumeralCaps{2}]:H$\alpha$ ratios are characteristic of shocked gas \citep{long2010}.

The primary focus of this paper is to constrain the progenitors of CCSNe, not SN Type Ia. Even though these catalogs do not provide the type of SN that created each SNR, there are ways to reduce the SN~Ia contamination in the catalogs. The SN~Ia rate is about one-fourth of the overall SNe rate  \citep{li2011}, and for M33 the SN~Ia fraction is expected to be less than for a galaxy like M31. CCSNe are associated with the explosion of massive stars
\citep{smarttetal2009}, and therefore younger stellar populations. While Type Ia SNe are associated with older stellar
populations. Therefore, by eliminating any SNR with zero SF within the last 80 Myr one can
effectively remove likely Type Ia SNRs from the analysis. \citet{jennings2014} took this approach and found that the fraction of SNRs with no SF in the last 80 Myr was consistent with the fraction of expected SN~Ia in M31 and M33.  If there is any SN~Ia contamination in our catalogs, the fraction will be low and will not have a statistically significant impact on the distribution. 
\\

\subsection{Star Formation Histories}
\label{section:SFH}

The SFHs that we use originate from (\citet{jennings2014}; for the SNRs in M33) and from (\citet{lewis2015}; for the M31 SNRs). To infer the SFHs, these authors first calculate the
photometry for all stars surrounding an SNR at a given distance.
\citet{jennings2014} used DOLPHOT \footnote{The
	original reference is \citet{dolphin2002}, and updated versions are
	available online.} to calculate the photometry of all stars within 50 pc
of the SNRs in M33. Later,  \citet{lewis2015} calculated the SFH for M31 in 100
pc $\times$ 100 pc regions throughout the Panchromatic Hubble Archive
Treasury (PHAT) footprint \citep{williams2014b}.  For each SNR in the \citet{leeandlee2014}
M31 catalog, we use the SFH from the corresponding 100 pc $\times$ 100
pc region calculated by \citet{lewis2015}. They
selected stars with S/N~$>$~4, sharpness squared~$<$~0.15, and
crowding~$<$~1.3.  These parameters ensure that the objects are
high probability (high S/N), not extended sources (sharpness), and distinguishable from
neighboring stars in crowded fields (crowding). For the details of deriving the SFHs, we refer the reader to those manuscripts.

The authors then derive the SFHs from the color magnitude diagram (CMD) for each field using the program MATCH
\citep{dolphin2002,dolphin2012,dolphin2013}.  MATCH generates model
CMDs that include the effects of observational errors, foreground and
internal dust extinction, and distance, and then generates SFHs to
maximize the likelihood of the observed CMD.  The modeled magnitudes
and colors are based upon stellar evolution tracks and isochrones from
\citet{marigo2017}, which is an updated version of PARSEC \citet{girardi2010}.
See the respective
manuscripts for the extinction and distances used. 

A major assumption of this
technique is that the young population within $\sim$~50 pc is coeval
with the progenitor. Stellar cluster studies suggest that over 90\% of stars form in clusters containing more than 100 members with $M$ $>$ 50 $\msun$ \citep{lada2003}. Furthermore, these stars likely remain spatially correlated on physical scales up to  $\sim$100 pc during 100 Myr. This spatial correlation continues even for low mass clusters that are not gravitationally bound \citep{bastian2006}. Therefore, by studying the stars surrounding these SNRs, we can determine the age of the star that exploded. Our previous studies \citep{gogarten2009,murphy2011,williams2014} have confirmed that this assumption is reasonable.

 In both analyses, the
SFH is calculated using logarithmic spaced age bins ($\Delta
\log_{10}(t/\text{yr})$), and the youngest edge of the minimum age bin
is $\log_{10}(t/\text{yr}) = 6.6$. This technique is similar to isochrone fitting, but it uses the entire CMD to infer the recent SFH. Therefore, unlike simple isochrone fitting, this
technique fits for multiple ages. Figure~\ref{fig:data}, shows four
examples of the SFH derived by MATCH.  



Even though \citet{jennings2014} reported SFHs for the SNRs in both M31 and
M33, we only use their M33 SFHs.  The SNR catalogs
that they used for M31 lack homogeneous SNR identification:
\citet{magnier1995}, \citet{braunwalterbos1993}, and
\citep{williams1995}. They were mainly identified using
	  [S~II]-to-H$\alpha$ ratios and there was no confirmation using
	  more reliable techniques, such as radio or X-ray
	  observations. Later, \citet{leeandlee2014} published an M31 SNR
	  catalog with many more observations to constrain SNR
	  candidacy. This was the first full coverage catalog with a
	  homogeneous survey using [S~II]-to-H$\alpha$.  To identify SFHs
	  for the SNRs in M31, we cross-correlate the SNR positions from
	  \citet{leeandlee2014} with the spatially resolved catalog of
	  SFHs in the PHAT footprint \citep{lewis2015}.  The
	  cross-correlations yields 65 SNRs with at least some SFH in the
	  last 80 Myr.  Table~\ref{m31table}
	  gives the median ages and corresponding progenitor mass for each SNR in M31.

\startlongtable
\begin{deluxetable*}{ccccccccc}
	\tablecaption{Ages and progenitor initial masses for SNRs in M31 \label{m31table}}
	\tablehead{
		\colhead{SNR ID} & \colhead{R.A. (J2000.0)} & \colhead{Decl. (J2000.0)} & \colhead{Progenitor} & \colhead{$\sigma_+$} & \colhead{$\sigma_-$} & \colhead{$\mzams$}  & \colhead{$\sigma_+$} & \colhead{$\sigma_-$}  \\
		\colhead{} & \colhead{(Degree)} & \colhead{(Degree)} & \colhead{Age (Myr)} & \colhead{Age (Myr)} & \colhead{Age (Myr)} &  \colhead{($\msun$)} & \colhead{($\msun$)} & \colhead{($\msun$)}
	}
	\colnumbers
	\startdata
	8   & 010.0589018 & +40.620914  & 45.0 & 0.1  & 21.9 & 7.7  & 3.1 & 0.0  \\
	10  & 010.126564  & +40.721081  & 22.2 & 3.3  & 3.6  & 11.0 & 1.2 & 0.8  \\
	13  & 010.1393518 & +40.726681  & 7.9  & 8.8  & 0.0  & 23.3 & 0.1 & 10.3 \\
	14  & 010.1402712 & +40.546425  & 22.8 & 1.8  & 3.1  & 10.8 & 0.9 & 0.5  \\
	16  & 010.1652269 & +40.580177  & 18.6 & 1.4  & 1.6  & 12.2 & 0.7 & 0.5  \\
	32  & 010.3917561 & +41.247833  & 45.0 & 0.1  & 7.8  & 7.7  & 0.7 & 0.0  \\
	34  & 010.3986397 & +41.115501  & 42.6 & 0.0  & 9.4  & 7.9  & 0.9 & 0.0  \\
	37  & 010.5428724 & +40.86359   & 45.0 & 0.0  & 4.3  & 7.7  & 0.4 & 0.0  \\
	42  & 010.6060371 & +40.873493  & 35.7 & 1.0  & 2.9  & 8.6  & 0.3 & 0.1  \\
	45  & 010.6318827 & +41.101646  & 14.2 & 4.5  & 0.4  & 14.4 & 0.3 & 2.3  \\
	46  & 010.6862478 & +40.909912  & 28.4 & 0.7  & 2.3  & 9.6  & 0.5 & 0.1  \\
	47  & 010.6966658 & +41.022324  & 22.8 & 1.2  & 2.4  & 10.8 & 0.7 & 0.3  \\
	51  & 010.730814  & +40.996078  & 12.1 & 17.0 & 0.2  & 16.2 & 0.2 & 6.7  \\
	54  & 010.7659559 & +41.604427  & 45.0 & 0.7  & 3.2  & 7.7  & 0.3 & 0.1  \\
	59  & 010.7866192 & +41.05183   & 22.7 & 1.5  & 2.0  & 10.9 & 0.6 & 0.4  \\
	60  & 010.7957678 & +41.627312  & 23.0 & 0.5  & 3.0  & 10.8 & 0.9 & 0.1  \\
	62  & 010.8106556 & +40.909092  & 17.9 & 0.7  & 1.1  & 12.4 & 0.5 & 0.3  \\
	63 & 010.8312664 & +41.050423  & 45.0 & 0.1  & 19.2 & 7.7  & 2.4 & 0.0  \\
	64  & 010.8446941 & +41.109844  & 28.4 & 0.8  & 2.1  & 9.6  & 0.4 & 0.1  \\
	65  & 010.8452797 & +41.098709  & 21.3 & 1.1  & 7.8  & 11.3 & 3.7 & 0.3  \\
	68  & 010.8975859 & +41.235611  & 28.4 & 0.4  & 3.0  & 9.6  & 0.6 & 0.1  \\
	70  & 010.913332  & +41.448254  & 27.9 & 0.1  & 12.8 & 9.7  & 4.2 & 0.0  \\
	73  & 010.9388771 & +41.446407  & 24.0 & 1.7  & 2.3  & 10.5 & 0.6 & 0.4  \\
	75  & 010.9473724 & +41.214676  & 36.5 & 0.6  & 3.9  & 8.5  & 0.5 & 0.1  \\
	77  & 010.9744349 & +41.688019  & 17.9 & 1.3  & 1.6  & 12.4 & 0.8 & 0.5  \\
	82  & 011.0045404 & +41.351501  & 22.5 & 1.9  & 3.1  & 10.9 & 0.9 & 0.5  \\
	84  & 011.0212221 & +41.455238  & 16.3 & 1.4  & 7.4  & 13.2 & 7.4 & 0.7  \\
	85  & 011.0231409 & +41.336361  & 42.1 & 0.3  & 16.1 & 7.9  & 2.1 & 0.0  \\
	86  & 011.0551329 & +41.841881  & 42.7 & 0.0  & 24.1 & 7.9  & 4.3 & 0.0  \\
	93  & 011.1101227 & +41.816521  & 14.9 & 2.5  & 0.9  & 14.0 & 0.6 & 1.3  \\
	94  & 011.1169834 & +41.303799 & 45.0 & 0.1  & 5.0  & 7.7  & 0.4 & 0.0  \\
	95  & 011.1231985 & +41.878918  & 37.6 & 0.4  & 14.3 & 8.4  & 2.3 & 0.0  \\
	98  & 011.1525688 & +41.418209  & 9.0  & 5.7  & 0.0  & 20.5 & 0.0 & 6.4   \\
	100 & 011.1617689 & +41.424114  & 27.9 & 1.4  & 1.6  & 9.7  & 0.3 & 0.3  \\
	105 & 011.1915255 & +41.88311   & 23.8 & 1.4  & 2.3  & 10.6 & 0.6 & 0.3 \\  
	107 & 011.2065029 & +41.885338  & 35.7 & 0.9  & 3.1  & 8.6  & 0.4 & 0.1  \\
	109 & 011.2106609 & +41.906368  & 52.5 & 0.0  & 33.4 & 7.2  & 4.8 & 0.0  \\
	110 & 011.2119102 & +41.536724  & 12.8 & 2.6  & 0.8  & 15.6 & 0.8 & 1.8  \\
	113 & 011.2269049 & +41.5312    & 40.8 & 0.2  & 18.9 & 8.1  & 3.0 & 0.0  \\
	116 & 011.2566252 & +41.992016  & 22.5 & 1.2  & 2.3  & 10.9 & 0.7 & 0.3  \\
	117 & 011.2709312 & +41.648121  & 27.0 & 0.8  & 7.7  & 9.9  & 2.0 & 0.2  \\
	118 & 011.2815895 & +41.596478  & 42.5 & 0.0  & 21.5 & 7.9  & 3.4 & 0.0  \\
	119 & 011.2828627 & +41.539677  & 19.7 & 12.5 & 1.0  & 11.7 & 0.4 & 2.8  \\
	121 & 011.2954798 & +41.668118  & 43.1 & 0.0  & 21.2 & 7.9  & 3.2 & 0.0  \\
	122 & 011.3005543 & +41.838196  & 4.7  & 29.6 & 0.0  & 47.0 & 0.0 & 38.2 \\
	123 & 011.307682  & +41.596668  & 45.6 & 0.0  & 14.8 & 7.7  & 1.5 & 0.0  \\
	125 & 011.3134565 & +41.573505  & 29.3 & 0.6  & 2.8  & 9.4  & 0.5 & 0.1  \\
	131 & 011.3582821 & +41.722782  & 51.7 & 0.0  & 29.7 & 7.2  & 3.8 & 0.0  \\
	135 & 011.3659668 & +41.774178  & 29.4 & 1.4  & 2.3  & 9.4  & 0.4 & 0.2  \\
	136 & 011.3698902 & +41.775375  & 27.8 & 0.9  & 2.3  & 9.7  & 0.5 & 0.2  \\
	137 & 011.3733635 & +41.791794  & 31.1 & 0.3  & 5.6  & 9.1  & 1.0 & 0.1  \\
	138 & 011.383029  & +41.801659  & 35.2 & 0.9  & 3.0  & 8.6  & 0.4 & 0.1  \\
	139 & 011.3980255 & +41.968945  & 24.5 & 10.6 & 1.8  & 10.4 & 0.4 & 1.8  \\
	141 & 011.4016514 & +41.79921   & 35.9 & 0.9  & 3.0  & 8.5  & 0.3 & 0.1  \\
	142 & 011.4078999 & +41.839176  & 33.8 & 0.6  & 15.1 & 8.8  & 3.3 & 0.1  \\
	145 & 011.4854908 & +42.186268  & 4.7  & 8.7  & 0.0  & 47.0 & 0.0 & 32.0 \\
	146 & 011.5171375 & +41.836693  & 11.3 & 12.9 & 0.0  & 17.1 & 0.0 & 6.6  \\
	147 & 011.5838375 & +41.88327   & 45.0 & 0.1  & 28.4 & 7.7  & 5.4 & 0.0  \\
	149 & 011.6344233 & +41.995224  & 18.8 & 0.7  & 1.4  & 12.1 & 0.6 & 0.2  \\
	151 & 011.6407747 & +41.993465  & 18.8 & 0.6  & 1.5  & 12.1 & 0.6 & 0.2  \\
	153 & 011.6587248 & +42.187496  & 40.7 & 0.0  & 22.1 & 8.1  & 4.1 & 0.0  \\
	154 & 011.662818  & +42.12149   & 44.7 & 0.0  & 4.7  & 7.7  & 0.4 & 0.0 \\
	\enddata
	\vspace{1ex}
	
    \footnotesize \textbf{Note.}  These SNRs are from the
		\citet{leeandlee2014} catalog.  The SFHs used to derive these
		ages and masses are from \citet{lewis2015}.  Column (1) gives the SNR ID. Column (2) and (3) is the position of the SNR. Column (4) is the median age from the SFH. The uncertainties in the SFH allow for a range of median ages; Columns (5) and (6) give the 68\% percentiles on the median age.  Columns (7), (8), and (9) give the corresponding mass and uncertainties.
\end{deluxetable*}
\hfill

There are three primary sources of uncertainty in estimating the age for each SNR.  For one, the resolution of the SFH limits the certainty
for each age bin. The resolution of each age bin is $\Delta \rm{{log}}_{10}($$t/$Myr) = 0.1 for M31 and $\Delta \rm{log}_{10}($$t$/Myr) = 0.05 for M33. Second, there are often multiple bursts of SF (see
Figure~\ref{fig:data}).  These multiple bursts often dominate the uncertainty in
estimating the age for each SNR. Third, the SF rate for each bin has an uncertainty.  

The primary purpose of this paper is to develop a
hierarchical Bayesian model to handle the multiple bursts.  In doing so, we automatically consider the
resolution of each age bin. The third source of uncertainty requires
translating the SFH and its uncertainty into a probability
distribution for the age.  We leave this transformation for future
work \citep{murphy2018}. For now, we simply convert the best-fit SFH into a PDF for each SNR. 

\subsection{Age Probability Densities for Each SNR Progenitor}
\label{subsection:agedistribution}

The first step is to convert the SFH into a progenitor age
distribution function for each SNR, $P_k(t)$, where the index $k$
references each SNR.  This PDF has units
of $1/{\rm Myr}$.  We assume that the probability density is
proportional to the star formation rate (SFR), and the normalization $M_{\star}$ is
the total amount of stars formed in the last $T_{\rm max}$ Myr:
\begin{equation}
\label{eq:SFHtoPDF}
P_k(t) = \frac{{\rm SFR}(t)}{M_{\star}(T_{\rm max})} \, {\rm for}\, t <
T_{\rm max} \, .
\end{equation}
Single-star evolutionary models predict a $\minmass$ for core collapse around $\sim$8 $\msun$ \citep{woosley2002}, which
corresponds to a $\tmax$ of $\sim$45 Myr \citep{marigo2017}.  To
properly model and infer this $\tmax$, the PDF must include ages above this.  Otherwise, the inference
algorithm would just detect the artificial cutoff in the PDF.  On the
other hand, if $T_{\rm max}$ is too large, then one adds significant
uncertainty in the form of SFH that is
clearly too old.  For this manuscript, we adopt $T_{\rm max} = 80$ Myr. 

The discreet version of the PDF for SFH is
\begin{equation}
\label{eq:SFHtoPDFDiscrete}
P_k(i) = \frac{SFR(i)}{M_{\star}(T_{\rm max})} \, ,
\end{equation}
where each bin is indexed by $i$ and the set of bins associated with
each SFH is $\{ i \}$.  Given this discrete PDF, the probability of a
star being associated with bin $i$ is
\begin{equation}
\label{eq:probofburst}
P_{\rm SF}(i) = P_k(i) \cdot \Delta t_i \, .
\end{equation}

The best-fit SFH for SNe and SNRs often show distinct bursts of SF.  
In many cases, the SFH is simple, and there is one clear burst of SF (Figure~\ref{fig:data}a) for an SNR. However, this is not always the case.  Sometimes there is more than one burst of SF (Figures \ref{fig:data}(b)-(d)). A priori, it is
unclear which burst is associated with the SNR, and this represents a
significant source of uncertainty in our analysis.  Therefore, to
properly infer the underlying progenitor age distribution, one needs
to also model the unassociated bursts of SF. In the following
derivation, we consider the SF in each bin, $i$, as independent bursts
of SF. 



\begin{figure*}
	\centering
	\subfloat{
		\includegraphics[width=\textwidth]{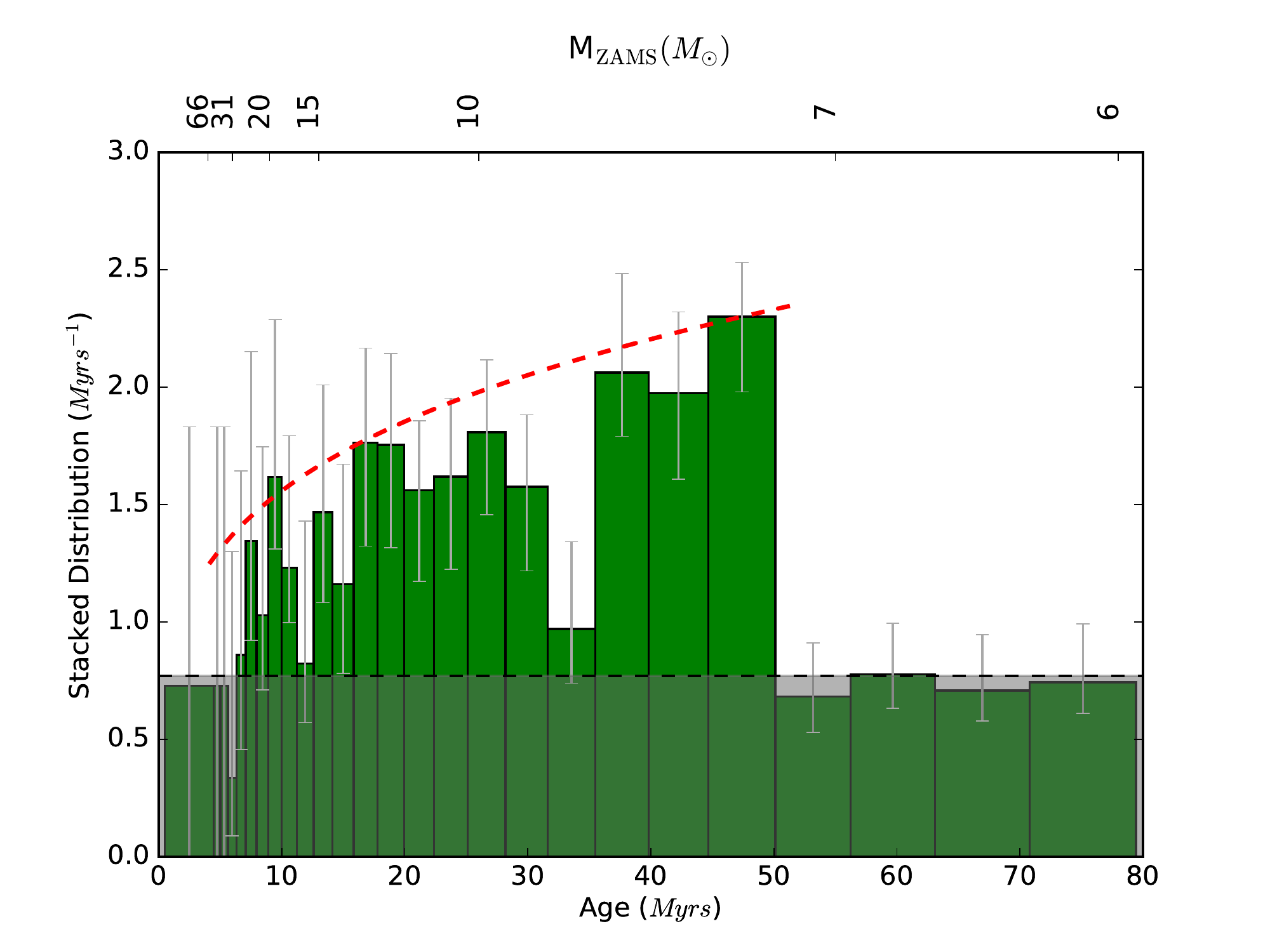}}
	\caption{The age distribution for
		all $\nsnr$ SNRs in M31 and M33 galaxies. For each SNR, we
		construct a probability distribution from the SFH; see Equation~(\ref{eq:SFHtoPDF}), and we add those distributions to get an age distribution for all SNRs in our sample. This stacked
		distribution, clearly shows two distributions. We model this as a
		power-law distribution with a
		$\tmin$, a $\tmax$, and a power-law slope (red dashed). There
		is also a contaminating distribution, a random uniform
		distribution that likely represents the random unassociated
		bursts of star formation (black dashed line).}
	\label{fig:stack_dist} 
\end{figure*}

\subsection{Progenitor Age Distribution Model}
\label{section:modelage}
 The simplest model that one might consider is a power-law distribution with minimum and maximum age.  This kind of model has the minimum number of parameters that one might expect for the distribution of SN progenitors.  To ensure that such a model is a reasonable approximation, we stack the age
distributions for all SNRs (see Figure~\ref{fig:stack_dist}).  In doing so, one can note that the SF
bursts seem to be drawn from two distributions.  One is the power-law
distribution associated with the SN progenitors.  The second is a
uniform distribution that probably represents random unassociated bursts.

Adding the PDFs for each SNR provides a reasonable approximation for
the overall progenitor age distribution.  This stacked distribution is a simple sum
of the individual PDFs
\begin{equation}
{\rm Stacked}(i) = \sum_k P_k(i) \, .
\end{equation}
 Figure~\ref{fig:stack_dist} shows the stacked age
distribution for all $\nsnr$ SNRs, and suggests a model for the age
distribution.
There are two clear components. There is an underlying
uniform distribution, $P_u(\hat{t})$, at all ages (black dashed line), which we presume is associated with the random unassociated
bursts of SF. We model this uniform distribution between 0
  Myr and $T_{\rm max} = 80$ Myr as $P_u(\hat{t}) = 1/T_{\rm max}$.
In addition, there is a power-law distribution with minimum and
maximum ages.  These minimum and maximum ages appear to be around
$\sim$~8~Myr and $\sim$~50~Myr. Therefore, we model the distribution of
true burst ages, $\hat{t}$, as a simple power-law distribution
	\begin{equation}
	\label{agepowerlaw}
	P_p(\hat{t}|\theta) \propto \hat{t}^\beta \cdot \Pi(\hat{t},\tmin,\tmax) \, ,
	\end{equation}
where $\Pi(\hat{t},\tmin, \tmax)$ is the unit boxcar function and is equal
to one between $\tmin$ and $\tmax$ and zero outside of this range. The model parameters are the $\tmin$, $\tmax$, and
slope of the distribution ($\beta$), which we collectively represent
as an array ($\theta$).  We propose that the true age of the SF burst
is either drawn from $P_p(\hat{t}|\theta)$ or $P_u(\hat{t})$, and the
true age, $\hat{t}$, for each burst represents a latent, or nuisance,
parameter in a hierarchical Bayesian model.

Formally, one would carry these nuisance parameters throughout the
derivation, making the derivation cumbersome until the very end.
If one assumes that the bins are not correlated, then one may
marginalize each bin to find the likelihood of drawing from
$P_p(\hat{t}|\theta)$ or $P_U(\hat{t})$.  The likelihood of bin $i$
having a burst from the uniform distribution is
\begin{equation}
\label{eq:puniform}
P_u(i) = \int_i P_u(\hat{t}) \,
d\hat{t} = \frac{\Delta t_i}{T_{\rm max}} \, ,
\end{equation}
The likelihood of bin $i$ having a burst from the power-law component is
\begin{equation}
P_p(i|\theta) = \int_i P_p(\hat{t}| \theta) \, d\hat{t} =
\frac{\tilde{t}_{\mathbin{i + \frac{1}{2}}}^{\beta+1} - \tilde{t}_{\mathbin{i - \frac{1}{2}}}^{\beta +
	1}}{\tmax^{\beta + 1} - \tmin^{\beta + 1}}\, .
\end{equation}
$\tilde{t}_{\mathbin{i + \frac{1}{2}}}$ and $\tilde{t}_{\mathbin{i - \frac{1}{2}}}$ are the
right and left sides of the bin unless the bin straddles either $\tmin$
or $\tmax$, the parameters for $P_p(i|\theta)$.  If the bin straddles
$\tmin$, then the left side is $\tmin$.  If the bin straddles $\tmax$,
then the right side is $\tmax$.

\begin{figure*}
	\centering
	\subfloat{
		\includegraphics[width=1.1\columnwidth]{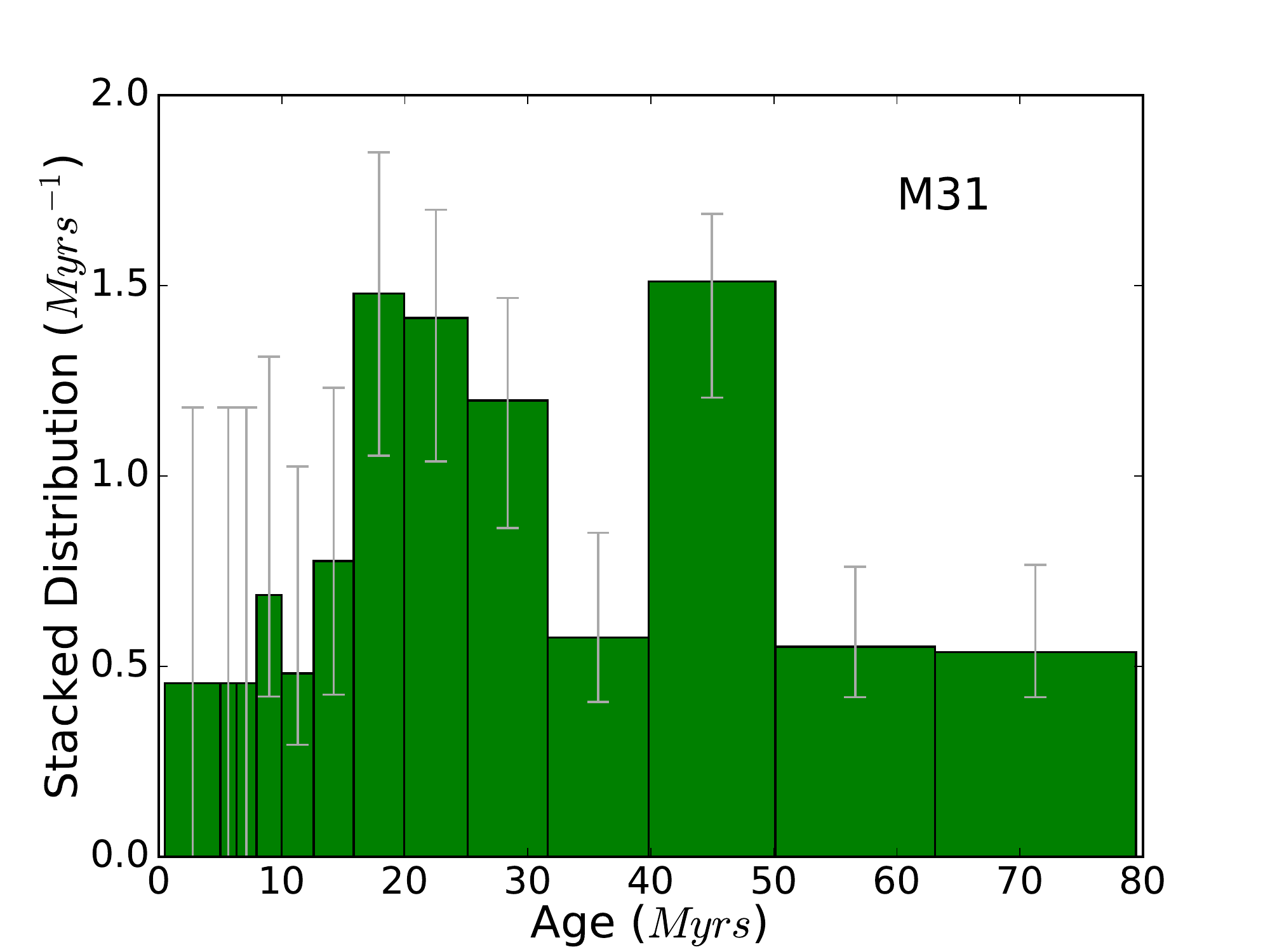}
	}
	\subfloat{
		\includegraphics[width=1.1\columnwidth]{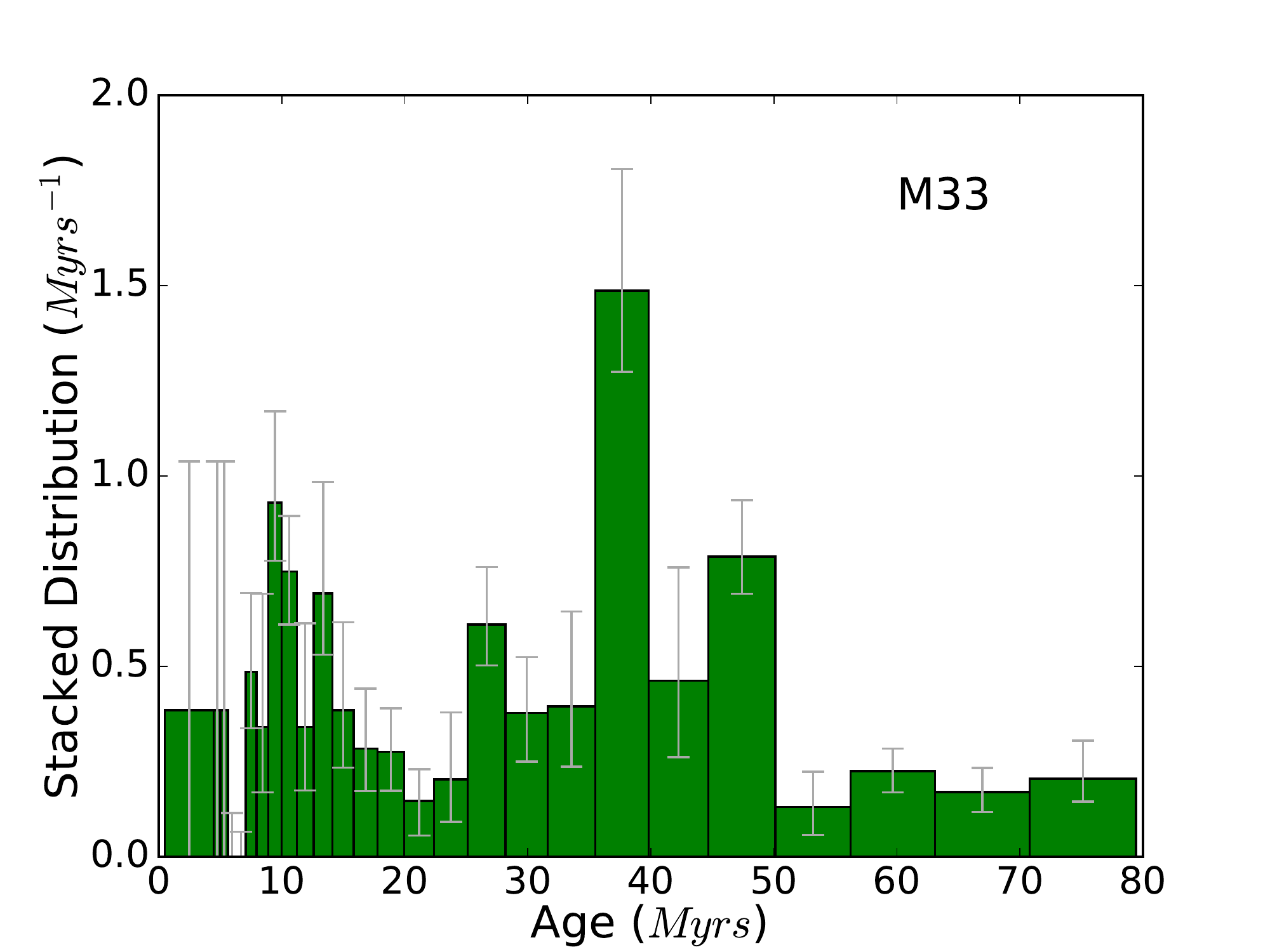}
	}
	\caption{The stacked age distributions for 62 SNRs in M31 and
		32 SNRs in M33. The samples for each galaxy do not have enough
		SNRs to clearly distinguish the progenitor distribution from the
		contamination. In fact, Bayesian inference of simulated data
		shows that it is difficult to constrain the distribution
		parameters (Figure~\ref{fig:params30}) with very few
		SNRs. Therefore, we emphasize the Bayesian analysis on the full data (Figure~\ref{fig:stack_dist}) rather than M31 or M33 alone.}
	\label{fig:stackm31m33} 
\end{figure*}

\begin{figure*}
	\centering
	\subfloat{
		\includegraphics[width=1\columnwidth]{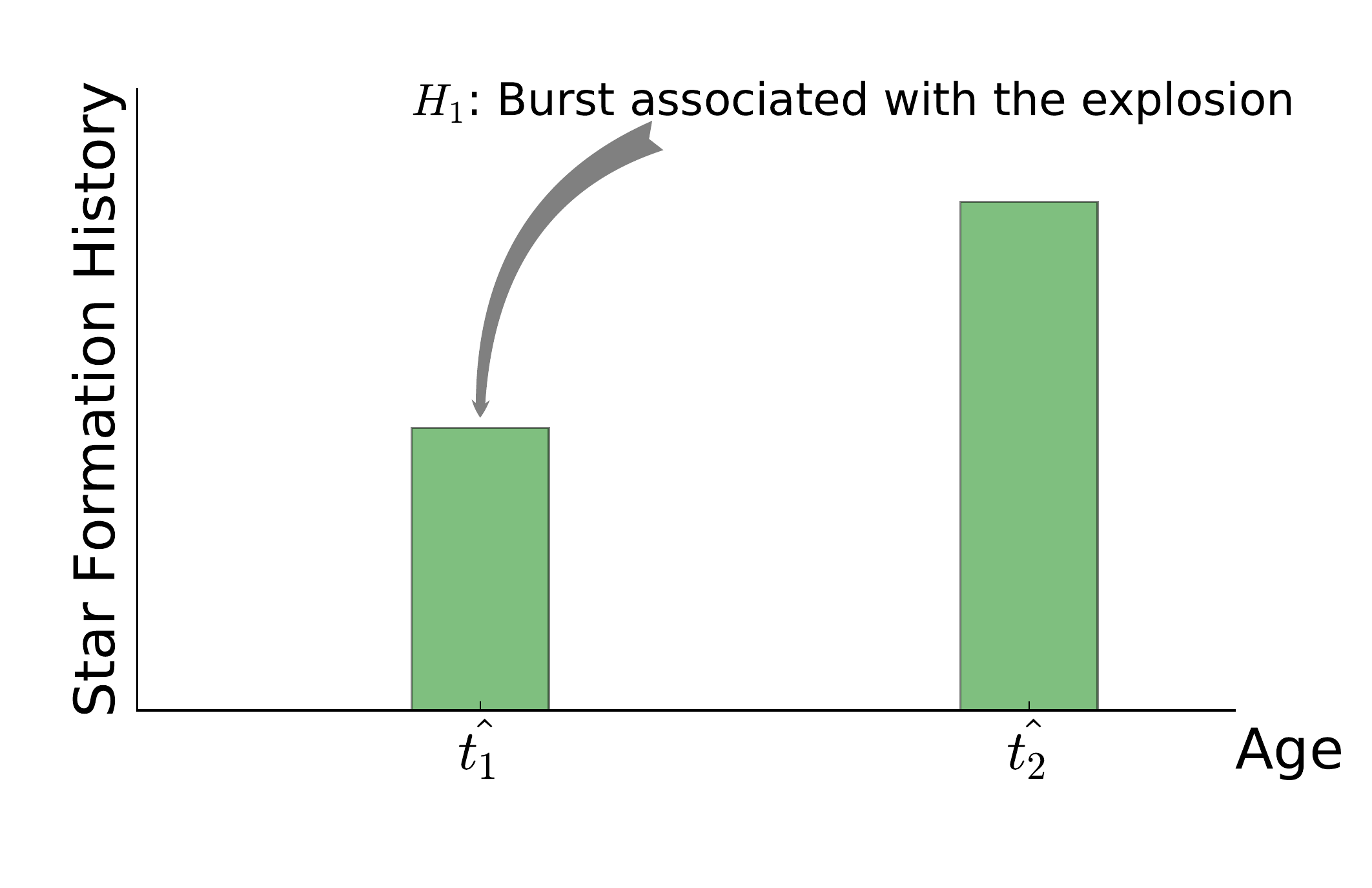}
		\label{fig:likeli1}}
	\subfloat{
		\includegraphics[width=1\columnwidth]{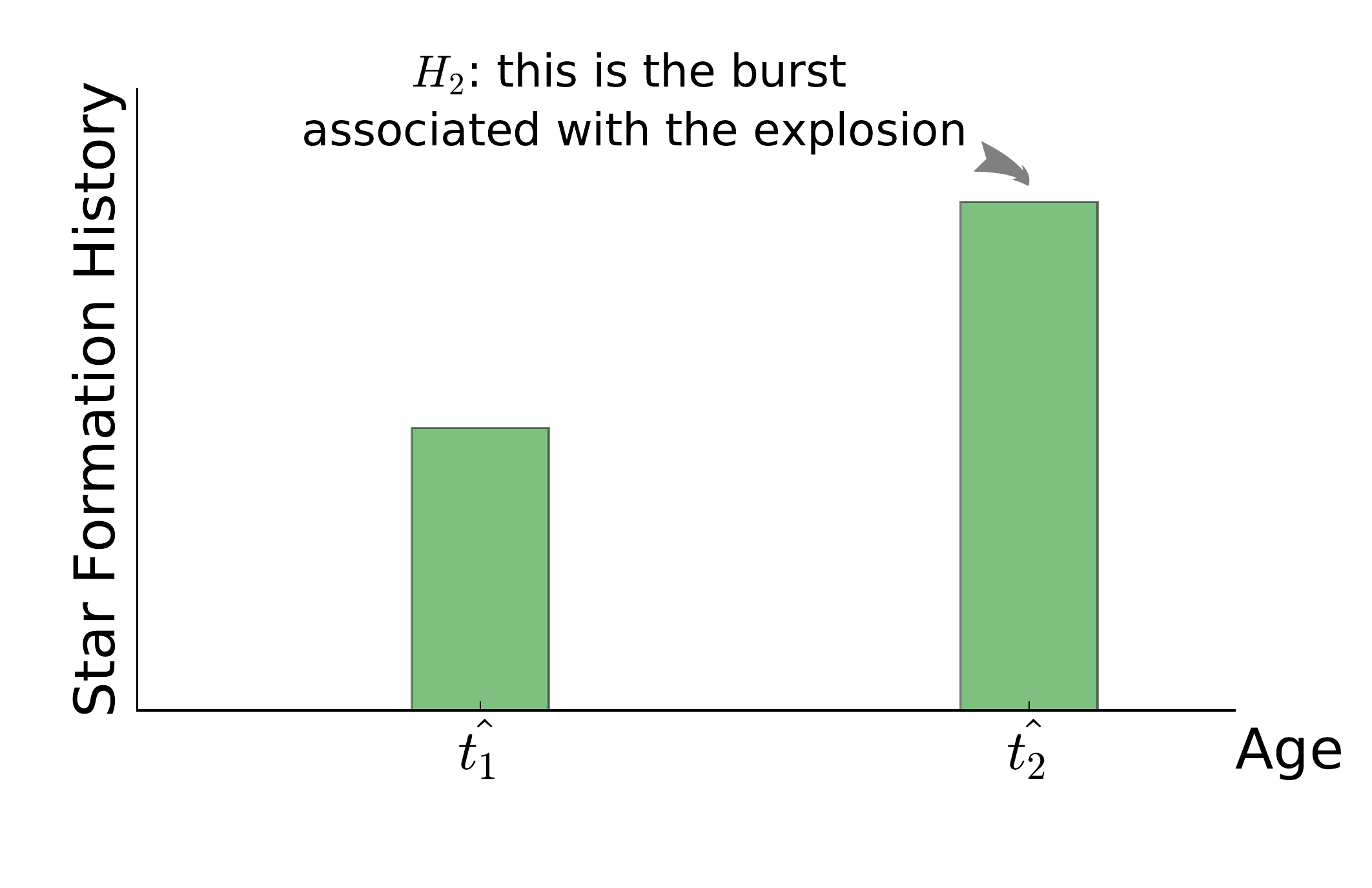}
		\label{fig:likeli2}}
	\caption{Likelihood model sketch for one SNR. We present two hypotheses ($H_1$, $H_2$), either the first peak is associated with the explosion ($H_1$), or the second peak is associated with the explosion ($H_2$). These two scenarios represent two different likelihood models, which we combined into just one likelihood model using an OR operation. The discrete random variable ``$H_i$" is the parameter that selects between those two scenarios. In the end, we marginalized over this nuisance parameter ``$H_i$".}
	\label{fig:likelimodel}
\end{figure*}

The simple age distribution model, which is quite apparent in the combined stacked distribution (see Figure~\ref{fig:stack_dist}) is not so clearly apparent for the individual galaxies. Figure~\ref{fig:stackm31m33} shows the stacked distributions for the individual galaxies, M31 (left panel) and M33 (right panel).  There are likely too few SNRs to clearly define the same model.  Therefore, this manuscript will focus on inferring the progenitor age distribution parameters for both galaxies.

One could model the aggregate stacked distribution in the Bayesian inference (Figure~\ref{fig:stack_dist}). However, there is more constraint in modeling each individual SNR. If one models only the aggregate, then one throws away additional constraints from modeling the likelihood of each individual SNR. For example, if one SNR has a very young age, then the likelihood of this one SNR will constrain the $\tmin$ to be quite small. Therefore, we choose to model the likelihood for each individual SNR.

\subsection{Hierarchical Bayesian Inference}
\label{sec:hierarchical}

To self-consistently infer all three
parameters, we use Bayes' theorem to compute the joint probability
$P(\theta|{\rm Data})$ of our model parameters, $\theta$, given the observations. The posterior distribution for each parameter is then the integral of the joint distribution over all other model parameters.  

Bayes' theorem states that 
\begin{equation}
\label{bayestheorem}
      P(\theta|{\rm Data}) = \frac{\mathcal{L}({\rm Data}|\theta) \ P(\theta)}{P({\rm Data})}
\end{equation}

The posterior distribution $P(\theta|{\rm Data})$ relates the probability of model parameters to the probability of observing the data, $\mathcal{L}({\rm Data}|\theta)$ (also known as likelihood),
and the prior distributions  $P(\theta)$. The prior distribution represents any prior knowledge one has about the parameters. $P({\rm Data})$ is the normalization.

In Bayesian inference, the primary task is to develop a likelihood
model for observing the data given the model parameters,
$\mathcal{L}({\rm Data}|\theta)$.  In this particular case, the data
consists of SFHs for each SNR.  If there are $N_{\rm SNR}$, then the
complete likelihood is the product of the likelihoods for each
individual SNR.
\begin{equation}
\label{eq:likelihoodofalldata}
\mathcal{L}({\rm data}|\theta) = \prod_k^{N_{\rm SNR}}
\mathcal{L}_k({\rm SFH}|\theta) \, .
\end{equation}
Each SFH is composed of a set of bins $\{ i \}$, which suggests a more specific
definition for the likelihood for each SNR: 
$\mathcal{L}_k({\rm SFH}|\theta) = \mathcal{L}_k(\{ i \}|\theta)$.

Now, we may derive the hierarchical likelihood model for observing one
burst that is drawn from $P_p(\hat{t}|\theta)$ and an arbitrary
number of random bursts drawn from $P_U(\hat{t})$.  For illustration
purposes, consider the case for which there are only two bursts, but it
is not known which is drawn from $P_p(\hat{t}|\theta)$ (see Figure~\ref{fig:likelimodel}).  In the following model, one and only one burst
is associated with the explosion; the other is a random unassociated
burst due to random uncorrelated star formation. One burst is labeled with 1, the other with 2.
We represent the parameters
of the power-law distribution by an array of parameters, 
$\theta = (\tmin, \tmax, \beta)$.  Our goal is to infer the posterior
distribution for these parameters.

When there are two bursts, and it is not clear which burst is drawn
from $P_p(i|\theta)$ or $P_u(i)$, then there are two hypotheses.  Hypothesis one ($H_1$) states that burst 1
is drawn from $P_p(1|\theta)$ and burst 2 is drawn from
$P_u(2)$.  Hypothesis two ($H_2$) states that burst 2
is drawn from $P_p(2|\theta)$ and burst 1 is drawn from
$P_u(1)$.  Since there is no a priori information on which hypothesis
is correct, $H_j$ represents a latent parameter of the hierarchical
model.Note that $j$ loops over each burst
	(or bin) just like $i$, with one important difference.  Hypothesis
$H_j$ represents the hypothesis when the SF in bin $j$ is assumed to be
	associated with the SN progenitor.

Now, we may derive a likelihood for each
SFH, $\mathcal{L}_k(\{ i \}|\theta)$.  But this likelihood depends upon
the latent parameters in $H$, so one must first define the joint
probability for the observed bursts {\it and} the latent
parameters,
$\mathcal{L}_k(\{ i \},H|\theta)$.   
Using the conditional probability theorem, the joint probability is 
\begin{equation}
\label{eq:conditional}
\mathcal{L}_k(\{i \},H_j|\theta) = \mathcal{L}_k(\{i\}|H_j,\theta) \cdot P(H_j) \, .
\end{equation}
Then to obtain the likelihood of
just the observed bursts, one marginalizes over the latent
parameter $H$:
\begin{equation}
\label{eq:marginalizelatent}
\mathcal{L}_k(\{ i\}|\theta) =
\sum_{j=1}^N \mathcal{L}_k(\{i \},H_j|\theta) \, ,
\end{equation}
where $N$ is the number of bins.

Next, we construct the hierarchical likelihood for each hypothesis.  For hypothesis $H_j$, the expansion of the
likelihood using the conditional probability theorem,
eq.~(\ref{eq:conditional}), becomes  
\begin{equation}
\label{eq:conditional2}
\mathcal{L}_k(\{i\},H_j|\theta) = P_p(j|\theta) \prod_{k\ne j} \cdot P_U(k) \cdot P(H_j) \, ,
\end{equation}
where $P(H_j)$ is the probability of the hypothesis. For hypothesis $H_j$, we set this to
the probability of a star being associated with the SF in bin $j$:
\begin{equation}
\label{eq:probHj}
P(H_j) = P_{\rm SF}(j) \, .
\end{equation}
Substituting these definitions, Equations~(\ref{eq:probHj}) and
(\ref{eq:conditional2}), into Equation~(\ref{eq:marginalizelatent}) leads to
the final form of the likelihood for SNR $k$, marginalized over all
latent parameters:
\begin{equation}
\label{eq:likelihoodfork}
\mathcal{L}_k(\{ i\}|\theta) =
\sum_{j=1}^N P_{\rm SF}(j) \cdot P_p(j|\theta) \prod_{k\ne j} \cdot
P_U(k) \, .
\end{equation}

Equation~(\ref{eq:likelihoodfork}) represents a general likelihood,
but one may further simplify this equation and reduce the
computational time for the calculation by reducing the number of
calculations within the MCMC runs.  The product series in
eq.~(\ref{eq:likelihoodfork}) is essentially only a function of bin
$j$, $f(j) = \prod_{k\ne j} P_U(k)$.  Since it is only a function of
$j$, this series may be calculated before the
MCMC runs.  In fact, with a little clever algebra, the likelihood
reduces even further.  If one multiplies and divides the
right-hand side of Equation~(\ref{eq:likelihoodfork}) by $P_U(j)$, then the
product series includes $k=j$ and the series is over all $k$ now,
i.e., $C = \prod_k^N P_U(k)$.  In other words, the product series is
now a constant and may be factored out of the summation.
Equation~(\ref{eq:likelihoodfork}) then becomes
\begin{equation}
\label{eq:likelihoodfork2}
\mathcal{L}_k(\{ i\}|\theta) =
C \cdot \sum_{j=1}^N \frac{P_{\rm SF}(j)}{P_U(j)} \cdot P_p(j|\theta) \, .
\end{equation}
Using the definitions of $P_{\rm SF}(j)$ in eq.~(\ref{eq:probofburst}) and
$P_U(j)$ in eq.~(\ref{eq:puniform}), the ratio $P_{\rm SF}(j)/P_U(j)$
becomes $P_k(j) \cdot T_{\rm max}$.  The final reduced likelihood is
\begin{equation}
\label{eq:likelihoodfork3}
\mathcal{L}_k(\{ i\}|\theta) =
C \cdot T_{\rm max} \cdot \sum_{j=1}^N P_k(j) \cdot P_p(j|\theta) \, ,
\end{equation}
and the product $C \cdot T_{\rm max}$ is simply a constant and may be
calculated before the MCMC runs.

With the likelihood for each SFH defined, one may now construct the
posterior distribution for $\theta$.  To find the likelihood for all
data, calculate the product of all likelihoods: insert
Equation~(\ref{eq:likelihoodfork}) or Equation~(\ref{eq:likelihoodfork3}) into Equation~(\ref{eq:likelihoodofalldata}).  Then the posterior
distribution for $\theta$ is proportional to this likelihood times
the priors for $\theta$:
\begin{equation}
\label{eq:posterior}
P(\theta|{\rm Data}) \propto \mathcal{L}({\rm Data}|\theta) \cdot
P(\theta) \, ,
\end{equation}
where $P(\theta) = P(\tmin) \cdot P(\tmax) \cdot P(\beta)$.

The priors, $P(\theta)$, for all model parameters are uniform with additional conditions specified in Table~\ref{tab:priortable}. Therefore, the model in eq.~\ref{eq:posterior} has only three unknown parameters
$\tmin$, $\tmax$, and $\beta$ embedded in $\mathcal{L}({\rm
  Data}|\theta)$. To infer the posterior distribution, $P(\theta|{\rm
  Data})$, we use the Markov Chain Monte Carlo (MCMC) sampler
\verb|emcee|, a python implementation \citep{foreman-mackey2013} of
the affine invariant ensemble sampler by
\citet{goodmanweare2010}. Typically, we use 10 walkers, 10,000 steps
each, and we burn 5000 of those.  Generally, the acceptance fraction
for an inference run is typically around $\alpha = 0.6$

\begin{table}
	\caption{Additional conditions for the priors.}
	\label{tab:priortable}
	\begin{tabular}{lc}
		\toprule
		Parameter  & Prior\\
		\midrule
		Minimum age $\tmin$ & $\mathcal{U}(0.5\, {\rm Myr},\tmax)$\footnote{Formally, this should go to zero, but we are considering power-law
			age distributions with a negative slope, so numerically we avoid the
			$\tmin = 0$.}\\
		Maximum age $\tmax$ & $\mathcal{U}(0.5\, {\rm Myr},T_{\rm max})$\footnote{ We analyze only the SNRs with SF within the last $T_{\rm max} = 80$ Myr.} \\
		Slope $\beta$ &  $\mathcal{U}(-1,10)$ \\
		\bottomrule
	\end{tabular}
\end{table}



\subsection{Testing the Hierarchical Model Using Simulated Data}
\label{sec:fakedata}

To test the above hierarchical model, we produce simulated normalized SFHs
for $N_{\rm SNRs}$.  For each test run, we set the number of SNRs to either 30, 100, and 300 SNRs.  For each SNR, one SF
event is drawn from $P_p(\hat{t})$ and $N_{\rm random}$
bursts from $P_U(\hat{t})$. However, we draw first $N_{\rm random}$ from
the Poisson distribution with a mean of $\lambda =1$.  We then map
these bursts into an SFH with the same resolution that we use in MATCH
runs $\Delta (\rm{log} \ t) = 0.05$.  For this simple test, the probability of
each burst is evenly split, $P_{\rm SF}(i) = 1/(N_{\rm bursts})$.

\begin{figure*}
	\centering
	\subfloat{
		\includegraphics[width=0.7\columnwidth]{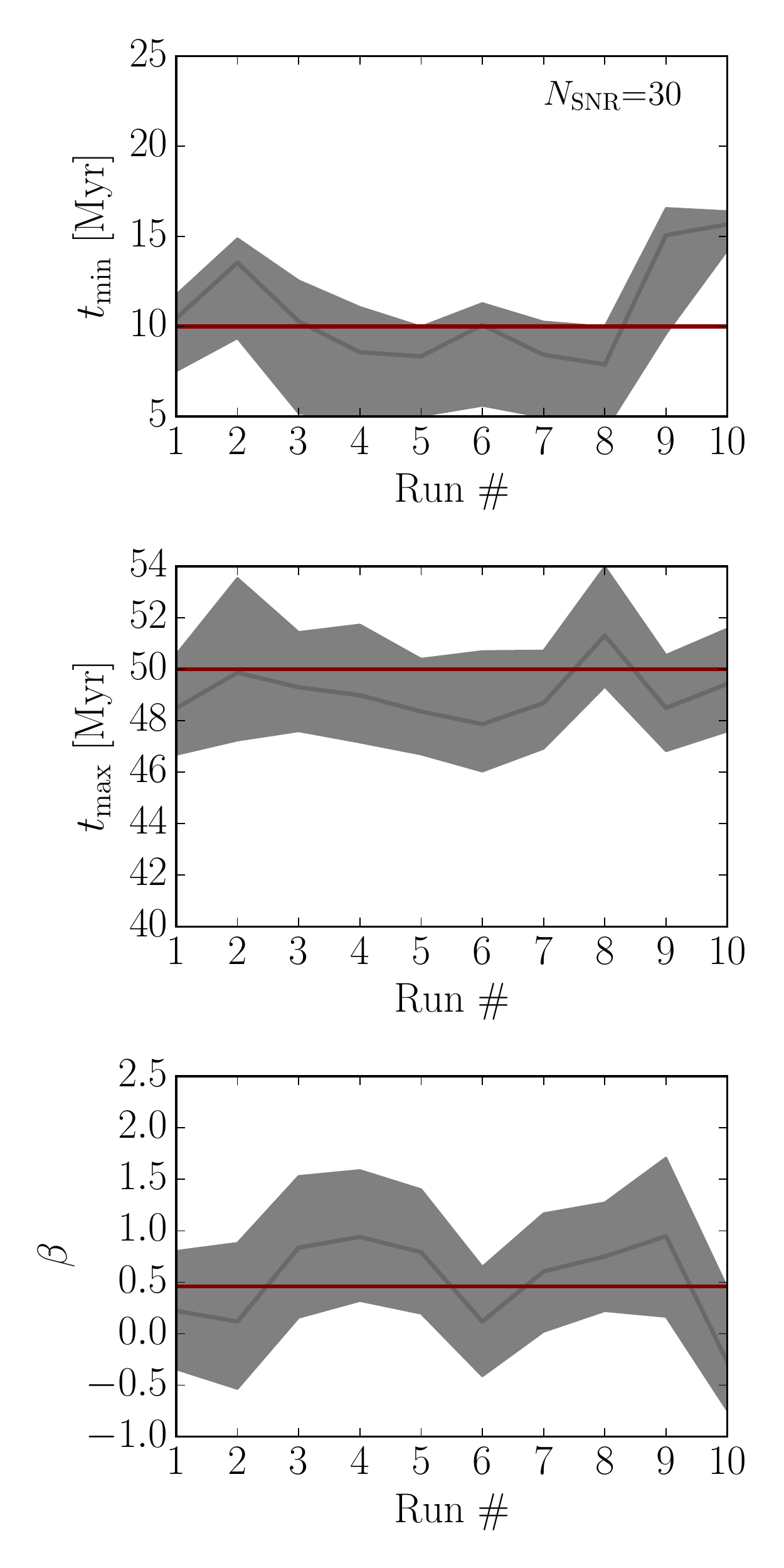}
		\label{fig:params30}}
	\subfloat{
		\includegraphics[width=0.7\columnwidth]{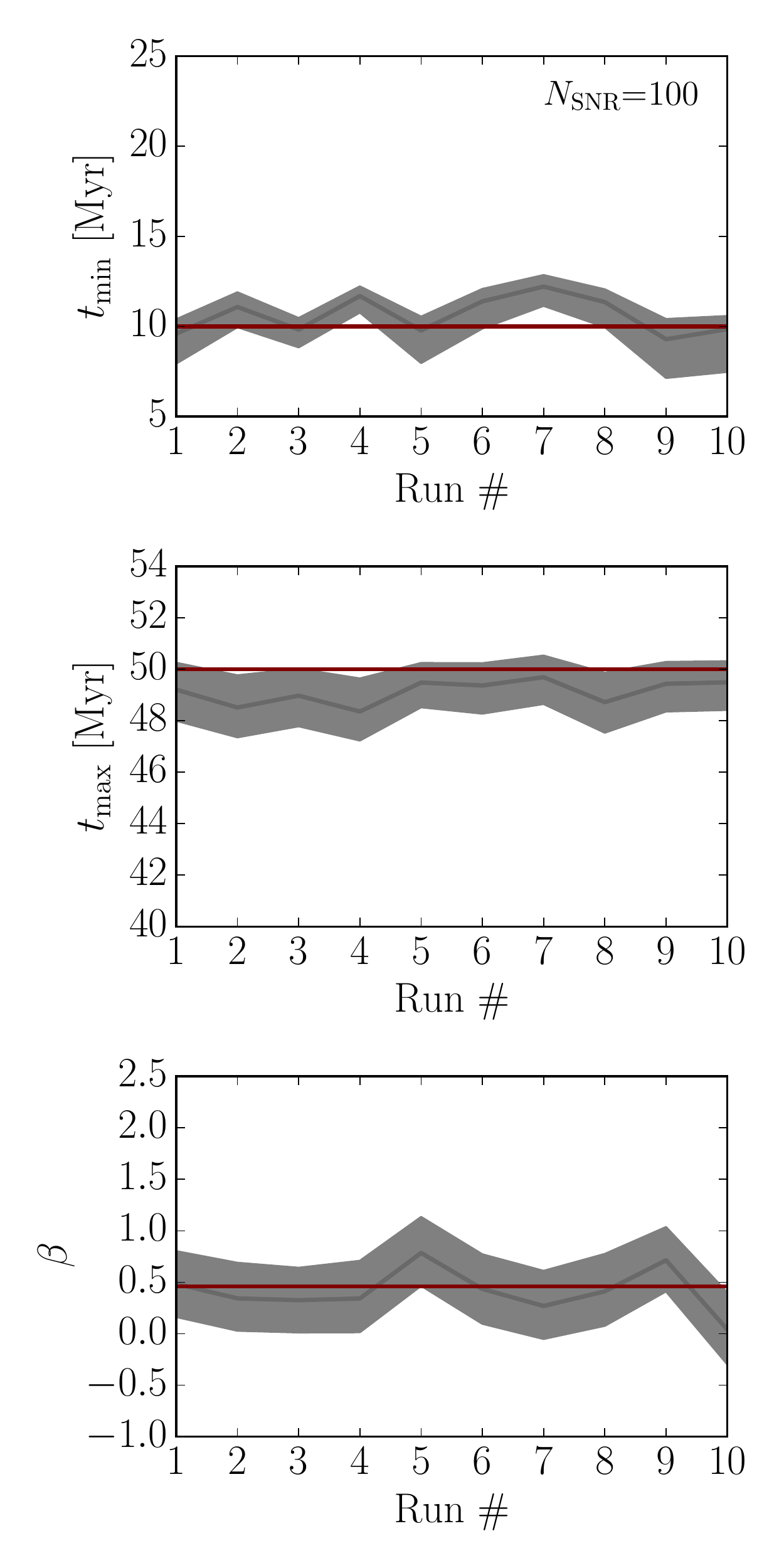}
		\label{fig:params100}}
	\subfloat{
		\includegraphics[width=0.7\columnwidth]{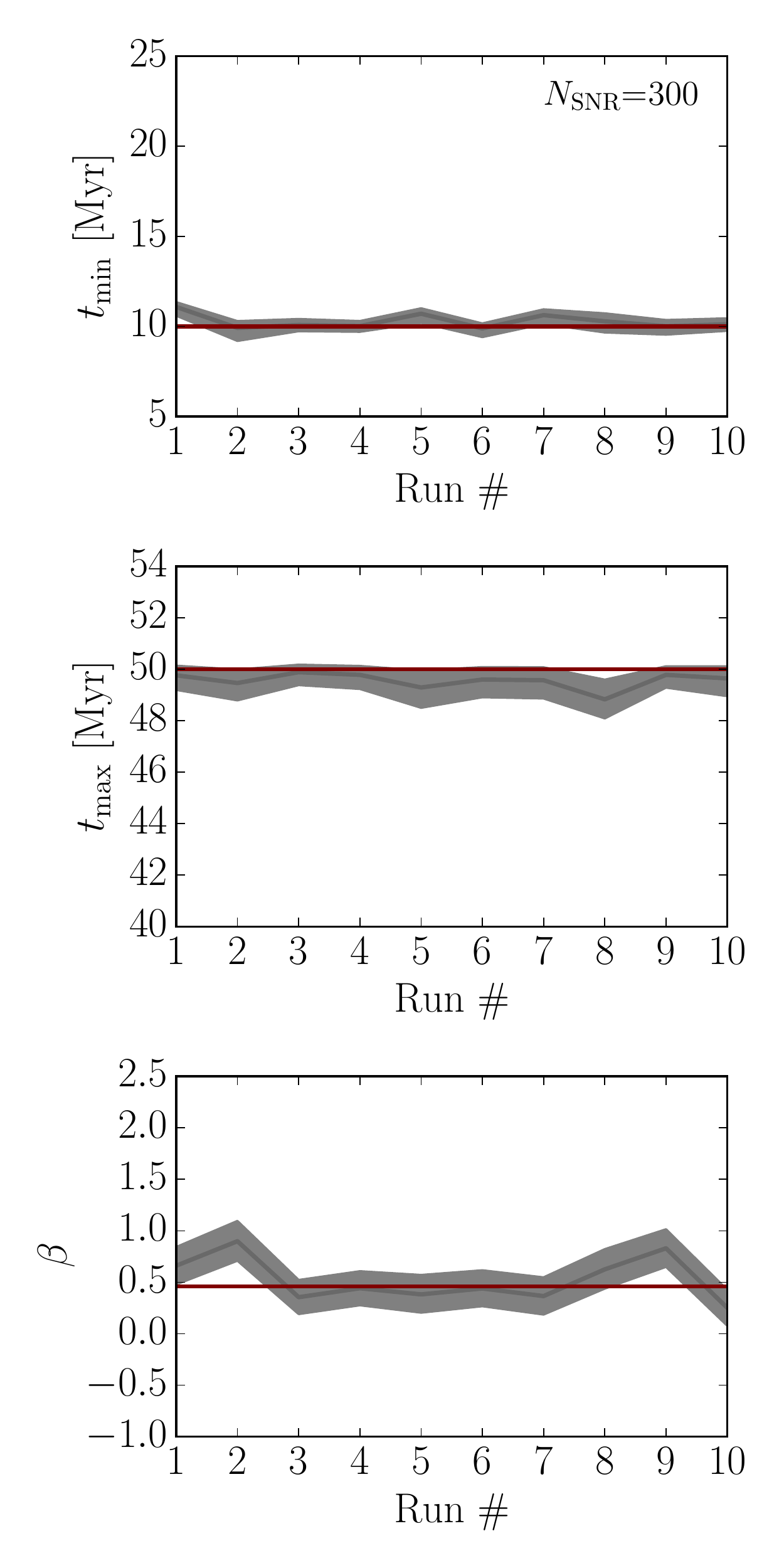}
		\label{fig:params200}}
	\caption{Marginalized parameters from simulated tests of the
	  Bayesian hierarchical model. From left to right, the number of
	  SNRs in each column is 30, 100, and 300. We run the test 10
	  times to compare the true parameter values (maroon solid
		line) with the median value of the marginalized posterior
	  distributions (gray solid line). If there is a bias in
	  any of the parameters, it is smaller or on the order of the statistical
	  uncertainties in every case.  These tests also show that the
	  progenitor age distributions are very poorly constrained for low
	  SNR numbers.}
	\label{fig:params}
\end{figure*}

To adequately test the method and to identify any biases, we run the
test 10 times with known parameters of $\tmin~=~10$~Myr, 
$\tmax~=~50$~Myr, and $\beta = 0.46$.  Figure~\ref{fig:params} show the
results when $N_{\rm SNR} = 30$, 100, and 300.  The horizontal maroon lines show the true parameter values.
The solid gray line shows the median value of the marginalized
posterior distributions and the gray band shows the 68\% confidence
interval.  With only $N_{\rm SNR} = 30$ the resulting marginalized
distributions are quite broad, providing very little constraint on the
parameters.  Note that there are only $\nsnrmto$ SNRs for M31 and
$\nsnrmtt$ for M33.  Since there are so few SNRs in each sample, we
emphasize the results from the combined data set.  

To calculate the bias, we calculate the
mean median value in each case, subtract this mean from the true value and report the uncertainty of the
median.  Neither $\tmin$ nor $\beta$ show any discernible systematic.
For $\tmax$ the potential systematics are 
$1.1 \pm 0.9$ for $N_{\rm SNR} = 30$, 
$0.7 \pm  0.4$ for $N_{\rm SNR} = 100$, and
$0.4 \pm 0.3$ for $N_{\rm SNR} = 300$.  We find that the bias is on the
order of the uncertainty in measuring the bias, and it gets smaller as
the certainty on the measurement increases (or the number of SNRs
increases).  Therefore, we suggest that there is no discernible bias.
If there is one, then it is significantly smaller than the bin width
at 50 Myr ($\Delta t = 5.8$ Myr).


\section{Results}
\label{section:results}

\begin{figure}
	\includegraphics[width=\columnwidth]{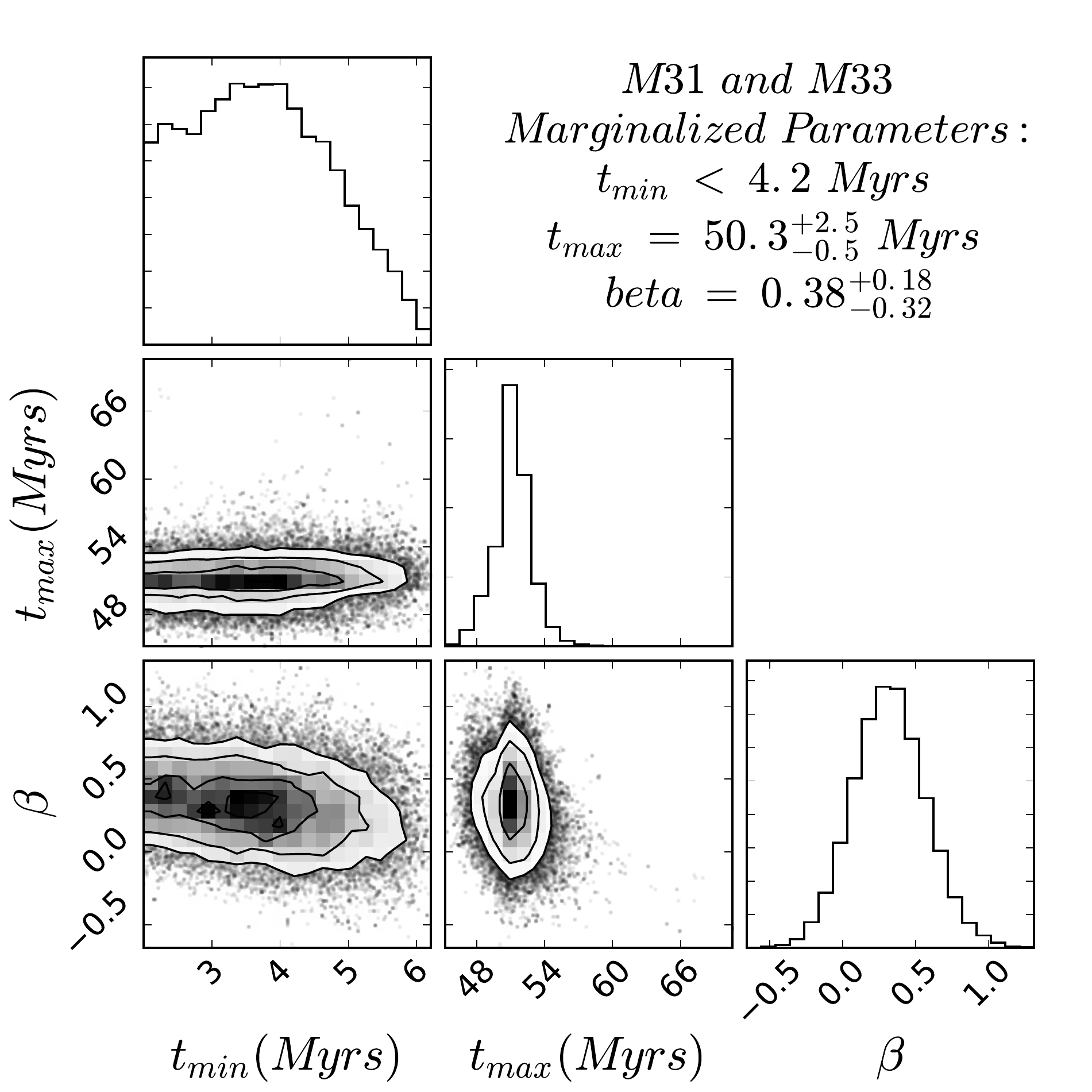}
	\caption {Posterior distributions for the model
	  parameters. We report the one-side 68\% confidence
		interval for the upper limit on $\tmin$, and the lower limit
		on $\minmass$.  For the rest of the parameters, we report the
		mode and the narrowest 68\% confidence interval for the
		uncertainty. Given the model shown in 	Figure
	  \ref{fig:likelimodel}, the minimum age is $\tmin < \minage$
	  Myr, the maximum age is  $\tmax = \maxage$ Myr, 	and the power-law slope is $\beta$ = $\betavalue$. The $\tmin$ is consistent with the minimum age for which MATCH can derive a star formation rate. 	Therefore, our $\tmin$ is actually an upper limit on the minimum age.}
	\label{fig:M31M33bayesages}
\end{figure}

\begin{figure}
	\includegraphics[width=\columnwidth]{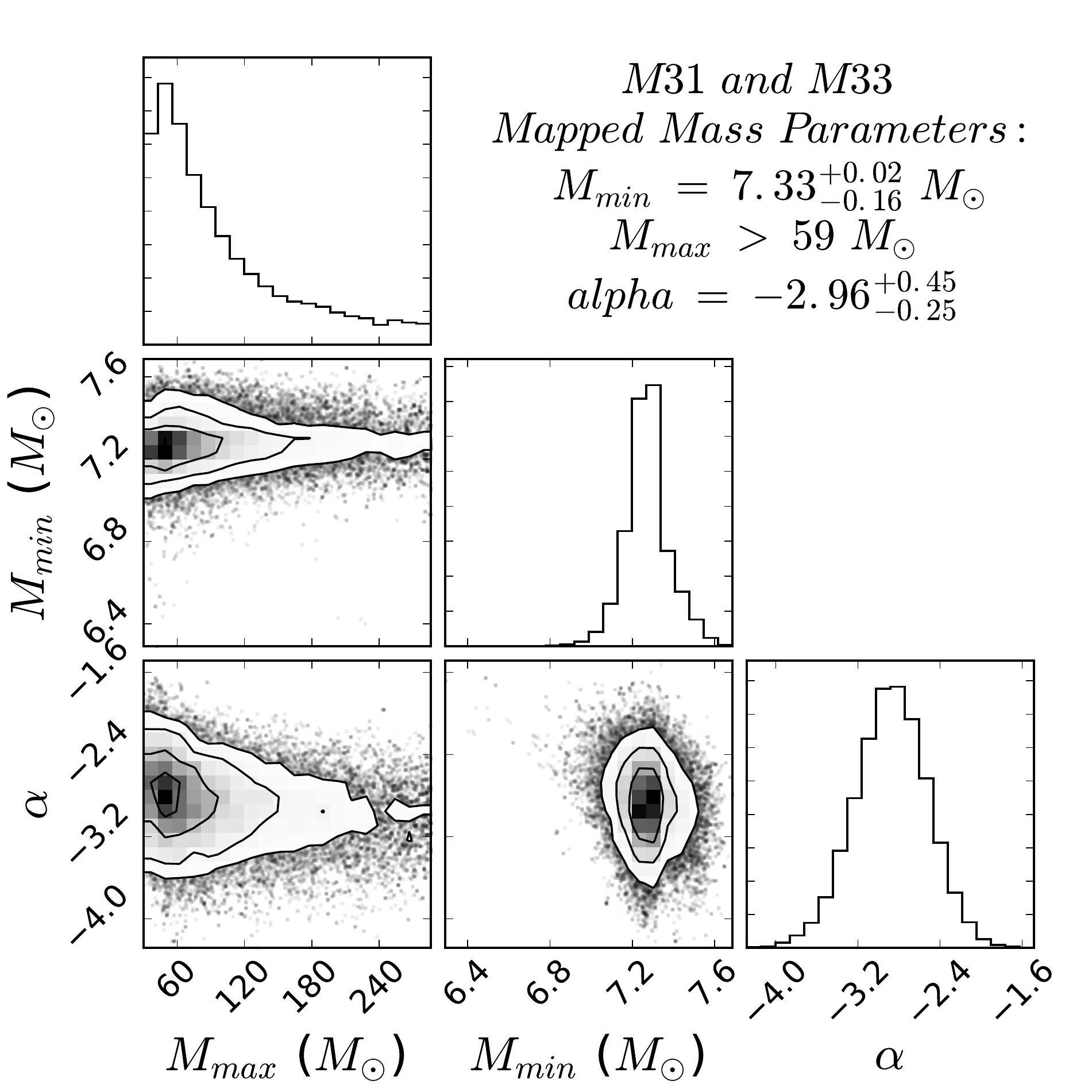}
	\caption{Posterior distribution for progenitor mass distribution
		parameters. To find $\minmass$, $\maxmass$, and
		the power-law slope in between, we use the results of
		Figure~\ref{fig:M31M33bayesages} and an age-to-mass mapping. The $\minmass$ is $\minmass =
		\minmassvalue$~$\msun$, the $\maxmass$ is $\maxmass >
		\maxmassvalue$ $\msun$, and the slope is $\alpha = \alphavalue$,
		 somewhat
		  steeper than the Salpeter initial mass function
		($\alpha$ = $2.35$). The $\maxmass$ is consistent
		with the detection limit at which MATCH can derive a star formation
		rate. Therefore, the $\maxmass$ is
		actually a lower limit on the $\maxmass$.}
	\label{fig:M31M33bayesmasses}
\end{figure}

Figures~\ref{fig:M31M33bayesages} and \ref{fig:M31M33bayesmasses} show SNR results of the MCMC sampler when applying our model to the ages of both M31 and M33 SNRs. Figure~\ref{fig:M31M33bayesages} represents the primary
inference: the posterior for the $\tmin$, $\tmax$, and power-law slope $\beta$ for the age distribution. Then, to obtain
Figure~\ref{fig:M31M33bayesmasses}, we use an age-to-$\mzams$ mapping to recast this as posterior
distributions for the $\minmass$, $\maxmass$, and slope $\alpha$ in the mass distribution.

The marginalized values for the age distribution parameters are as
follows. The $\tmin$ is $\tmin < \minage$ Myr, the $\tmax$ is
$\tmax = \maxage$ Myr, and the power-law slope for the age
distribution is $\beta = \betavalue$. We use a one-sided 68\%
  confidence interval to
calculate the upper limit on $\tmin$, and we calculate the narrowest 68\% confidence interval for $\tmax$ and $\beta$. This upper bound on $\tmin$ is
roughly consistent with the upper end of the youngest age bin in MATCH
(4.47 Myr for M33 and 5.01 Myr for M31).  This age is an upper bound because all stars more massive
than about $\sim$60~$\msun$ start to have Eddington factors near one
(ratio of photon force compared to gravitational force).  When the
Eddington factor approaches one, $L \propto M$, and the lifetime,
$t_{\rm lifetime} \sim M c^2/L$,
becomes a constant for these stars.  Hence, all stars with masses 
$\gtrsim 60 \msun$ have lifetimes $\sim$4 Myr. 

The consequence for
age-dating is that it is
impossible to distinguish the ages of SF bursts that are younger than
$\sim$4 Myr; MATCH places all of these young bursts into the
youngest age bin, which ranges from 3.98 to 4.46 Myr for the M33
resolution and 3.98 to 5.01 Myr for the M31 resolution. Since the SFR in this youngest bin actually represents the SF between 0
and $\sim$5 Myr, we redefine the youngest age bin to include all ages
below $\sim$5 Myr.  Practically, we cannot move the left side of this
bin to zero, because we are inferring the parameters of a power-law
age distribution.  To avoid the singularity imposed by this assumption,
we set the left side of the youngest bin to 0.5 Myr. 

Even though single-star evolution most naturally predicts the progenitor mass distribution, our primary inference is on the progenitor age distribution.  For one, the fundamental data for each SNR is the SFH.  Secondly, the clear mapping between age and mass is only valid for a restricted set of single-star evolutionary models. Binary
evolution may significantly complicate this mapping. Therefore, to accommodate future binary analyses, we first infer the progenitor age distribution.  In this manuscript, we consider the most straight forward case, single-star evolution. 

\begin{figure}
	\includegraphics[width=1.05\columnwidth]{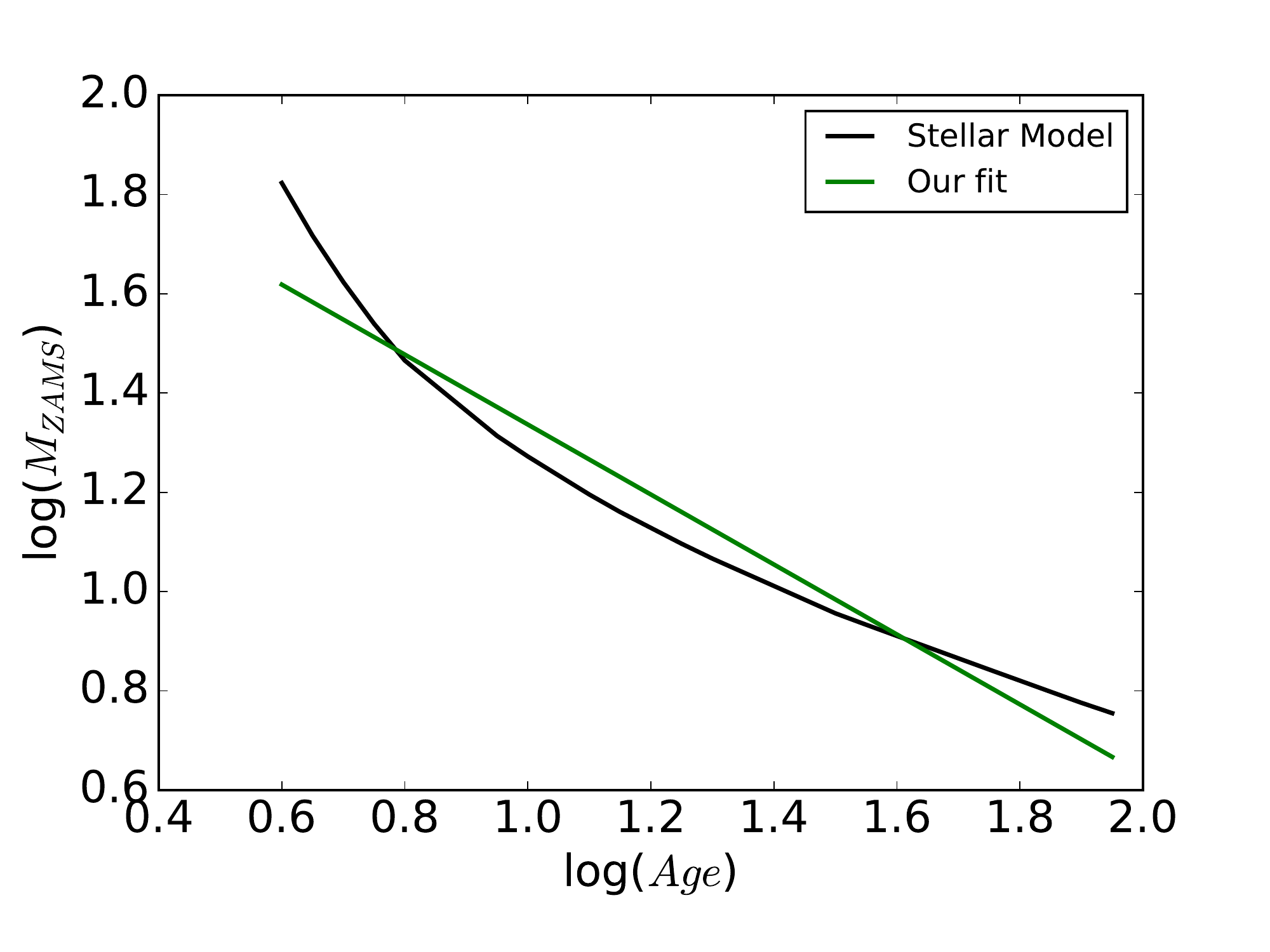}
	\caption{$\mzams$ vs. death time from stellar evolution models
		(black curve; \citealt{marigo2017}). We use this curve to map the
		progenitor age distribution parameters into the progenitor mass
		distribution parameters.  Specifically, we interpolate using
		the black curve to map $\tmin$ to $\maxmass$ and $\tmax$ to
		$\minmass$ (see section~\ref{section:results}). Mapping the
		power-law slope in age to a power-slope in mass is less
		trivial.  Therefore, we fit a log-linear line (green line) to the stellar
		evolution model (black curve) and use this fitting formula to map
		$\beta$ to $\alpha$.}
	\label{fig:agetomass}
\end{figure}

For this case, we map the progenitor age distribution to a progenitor
mass distribution. To do so, we make a few necessary assumptions. We
assume single-star evolution, solar metallicity of $Z$ = 0.019, and the
stellar evolution models of \citet{marigo2017}. To convert $\tmin$ and
$\tmax$ to its counterpart in mass space ($\maxmass$ and $\minmass$),
we use the results of stellar evolution models (solid, black curve in
Figure~\ref{fig:agetomass}). However, mapping the slope in age ($\beta$) to a slope in mass
(-$\alpha$) using the stellar evolution model curve is less trivial, and instead we use
a log-linear fit.  For a simple power-law age-to-mass mapping, the transformation would be analytic and simple. Unfortunately, the slope for the age-to-mass mapping from stellar evolution is not a single power-law slope.  Therefore, to determine the most appropriate power-law approximation, we fit a power-law to the age-to-mass mapping curve and use this to transform each $\beta$ in the
posterior distribution to an $\alpha$
	\begin{equation}
	  \label{eq:agetomass}
		m \propto t^{\frac{\beta+1}{1-\alpha}} =21.9 \ \msun \cdot \Big( \frac{t}{10 \rm {Myr}} \Big)^{-0.70}
	\end{equation}
Formally, the power-law index of the age-to-mass map changes slightly
from -1 to -0.6 in the mass range that we consider, but for the 
purposes of our simple mapping we use a log-linear fit that produces
a slope of -0.7
(see Figure~\ref{fig:agetomass}).

Finally, under the assumption that all CCSNe result from the explosion
of single stars, the posterior distributions for the mass distribution
parameters are in Figure~\ref{fig:M31M33bayesmasses}. The marginalized
parameters for all SNRs are $\minmass~=~\minmassvalue$~$\msun$, the
$\maxmass$ is greater than $\maxmass$~$>$~$\maxmassvalue$~$\msun$, and
the slope of the progenitor distribution is $\alpha$~=~$\alphavalue$.

Figure~\ref{fig:M31M33bayesmasses} emphasizes the results for both M31
and M33.  Table~\ref{tab:m31m33inference} reports the results for
both and for each galaxy separately.  As the simulated inferences
suggested, the low number of SNRs in each galaxy provides very loose
constraints on the parameters.  Hence, we emphasize the results
from the combined data set.
	
\begin{table*}
	\centering
	\caption{Inferred parameters for age and mass Distributions}
	\label{tab:m31m33inference}
	\begin{tabular}{l|cccccc}
		\toprule
		& $\tmin$ (Myr)  & $\tmax$ (Myr) & $\beta$ & $\minmass$ ($\msun$) & $\maxmass$ ($\msun$) & $\alpha$ \\
		\midrule
		M31 and M33 & $<$ $\minage$ & $\maxage$ & $\betavalue$ & $\minmassvalue$ & $>$ $\maxmassvalue$ & $\alphavalue$ \\
		M31 & $<$~$4.8$ & $54.3^{+13.9}_{-1.3}$ & $-0.05^{+0.34}_{-0.25}$ & $6.50^{+0.58}_{-0.17}$ & $>$~$46$ & $-2.35^{+0.36}_{-0.48}$ \\
		M33 & $<$~$9.7$ & $50.4^{+2.0}_{-1.9}$ & $1.07^{+0.63}_{-0.80}$ & $7.32^{+0.12}_{-0.14}$ & $>$~$19$ & $-3.94^{+1.13}_{-0.90}$\\
		\bottomrule
	\end{tabular}
	\vspace{1ex}
	
	\footnotesize \textbf{Note.}  We report one-sided 68\% confidence intervals
	for the upper limit on $\tmin$ and the lower limit on
	$\minmass$.  For the rest of the parameters, we report the mode
	and the the narrowest 68\% confidence interval. The main results are for the combined data-set
	(M31 and M33).  The constraints on the model parameters for the individual galaxies are very broad due to the low number of SNRs in each galaxy.
\end{table*}


\section{Discussion}
\label{section:discussion}

In general, the inferred $\minmass$, $\maxmass$, and slope $\alpha$ are consistent with previous estimates \citep{Smartt2004,smarttetal2009,jennings2014}.  The primary difference being that the uncertainties of this manuscript are more constrained.

\subsection{Minimum mass}
 
In general, stellar evolution theory predicts that stars above about
7-11 $\msun$ experience core collapse \citep{woosley2002}. This lower
mass limit depends on a variety of factors, such as the model selected and the
chemical composition of the star, e.g. helium abundance, metallicity,
convection, and  convective overshoot parameter
\citep{woosley2002}. For example, \citet{iben1983} reported that
a variation in chemical composition leads to a variation in the $\minmass$ of
$\minmass = 8-9 $ $\msun$. In addition to this, the overshoot
parameter can reduce the minimum value significantly. \citet{eldrigde2004} found that extra mixing, in the form of convective overshooting, moves the $\minmass$ to lower masses for SN. For example, in
\citet{bressan1993} they reported a value of $5-6$ $\msun$
  when assuming overshoot mixing by half of a pressure scale height
  for metallicity and abundance of $Z$ = 0.02 and $Y$ = 0.28.


Recent observational constraints on progenitor masses suggest that the
minimum mass is near the theoretical prediction. Based on the
observations of 20 type II-P SNe, the minimum value estimated for
CCSNe is $\minmassvalue$
\citep{smarttetal2009}. Using simple KS tests on a similar sample of SNRs
as in this manuscript, \citet{jennings2012} inferred a $\minmass$ of
7.3 $\msun$. These are all consistent with our more precise $\minmass$ determination of $\minmassvalue$ $\msun$.

In general, masses near the $\minmass$ are difficult to model, so
these results could provide new insight into late-stage evolution of
massive stars. However, we need to take binarity effects and biases
into account when modeling our data. The fact that we are getting a
$\minmass$ that is on the low side of possible predictions may be a sign that binarity plays a significant role in which stars explode. For example, in a binary, the primary star could explode giving a kick to the secondary star that then explodes in an older region. Mergers are another way in which a massive star could explode in a relatively old region. It is expected that nearly 24\% of massive stars merge and form rejuvenated stars \citep{sana2012}. The resulting merged star is more massive, but it will have an age that is more consistent with a lower mass star. This could cause associations between CCSNe and stellar populations that are otherwise too old to have single stars that would explode. \citet{zapartas2017} predict that $15^{+9}_{-8}$\% of CCSNe will be late explosions.
 
 \subsection{Maximum mass and slope of the distribution}
 
At the upper end of the progenitor age distribution, there are very few observational constraints. \citet{smarttetal2009} suggested that the upper mass for SN IIP progenitors is $\sim$ 17 $\msun$, which is significantly lower than the observed masses of RSGs, the progenitors of SN IIP.  However, their sample only included 18 detections and 27 upper limits on flux.  More recently, \citet{daviesbeasor2017} suggested that a more accurate application of bolometric corrections brings the upper limit mass up, more in line with the RSG observations. Even so, this upper limit for SN IIP progenitors does not need to be the upper limit for explosions in general. 

More massive stars are expected to lose much of their mass and explode
as other types of SNe. \citet{jennings2014} used KS statistics to
constrain the upper end of the progenitor mass distribution for 115
SNRs in M31 and M33. The KS statistic does not allow for a self-consistent inference of all distribution parameters.  Therefore, they
first estimated the $\minmass$, then estimated the slope assuming no
$\maxmass$, and found a slope of $\alpha = 4.2^{+0.3}_{-0.3}$.  They
also considered a second scenario; they set the slope to Salpeter ($\alpha = 2.35$) and inferred a $\maxmass$ of $\maxmass = 35^{+5}_{-4}$. In either scenario, they concluded that either the most massive stars are not exploding at the same rate as lower mass stars or there is a bias against observing SNRs in the youngest SF regions.

With a more sophisticated Bayesian analysis, in which we infer all
parameters simultaneously, we find that the $\maxmass$ is above
$\maxmassvalue$ $\msun$, and the slope is $\alpha$ = $\alphavalue$. The lower
limit on the $\maxmass$ is consistent with our detection
limit. Therefore, within our detection limits, we find no discernible $\maxmass$. The slope is somewhat steeper than Salpeter,
similar to \citet{jennings2014}, but not quite as steep.  This
significant difference demonstrates the advantage of inferring all parameters
of the progenitor distributions simultaneously.
 

 If there is a bias associated with the SNR catalog it
 might affect the $\maxmass$ and/or the slope of the distribution, making it steeper or more
 positive, but the bias is very
 unlikely to affect the $\minmass$. Observable SNRs are very likely biased to certain regions of
 the ISM and may be biased to certain ages
 \citep{elwood2017,sarbadhicary2017}.  We do not account for this
 bias, but instead report the progenitor age distribution and mass
 distribution as observed. However, as we incorporate the uncertainties in the
   SFH (future work), the distribution of minimum ages will likely
   incorporate even older ages.  With regard to the
   $\minmass$, Figure~\ref{fig:stack_dist} shows an abrupt drop at
   around 50 Myr.  It is very unlikely that a general environmental
   bias would mimic such a drop in the progenitor distribution.
   
    To improve the accuracy of these constraints, we identify several
    assumptions of our analysis that need verification or
    improvement. There are three dominant sources of uncertainty in the
    analysis: SFH resolution, multiple bursts of SF, and the uncertainty
    in the SFR. This current paper addresses the first two, but does not
    address the uncertainty in the SFR. In the future, we plan to develop
    a hierarchical likelihood model that includes the distribution of SFR
    for each age bin. For example, see \citet{murphy2018}. Furthermore, the nonlinear conversion from an age distribution,
    $P(t_{\rm age})$,
    to a mass distribution, $P(M)$, imposes a non-uniform prior on the mass
    distribution parameters. Explicitly, 
    $P(M) = P(t_{\rm age})\cdot dt_{\rm age}/dM $, and eq.~(\ref{eq:agetomass})
    implies that $dt_{\rm age}/dM$ is roughly 
    $t_{\rm age} \propto M^{-2.4}$.  When converting from $P(t_{\rm age})$
    to $P(M)$, $dt/dM$ acts like a prior.   In this particular case,
    $dt_{\rm age}/dM$ may impose a bias
    toward low masses for $\minmass$ and $\maxmass$.

\section{Conclusion}
\label{section:conclusion}
Using a Bayesian hierarchical model, we infer the progenitor age
distribution from the SFHs near $\nsnr$ SNRs. Of these SFHs, $\nsnrmtt$ are for
SNRs in M33, and $\nsnrmto$ are for SNRs in M31.  The SFHs for the $\nsnrmtt$ SNRs in
M33 were previously published in \citet{jennings2014}.  The remaining
$\nsnrmto$ SNRs in M31 are new and result from correlating the SNR locations
from \citet{leeandlee2014} with the resolved SFH map of M31 from
\citet{lewis2015}, Technically, there are 71 SNRs from
\citet{leeandlee2014} that are in the PHAT footprint, but 8 of these
had no SFH within the last 60 Myr.  These 8 (or 11\%) are most likely
SN Ia SNR candidates, which we omit from our catalog.  From the
remaining $\nsnr$ SNRs, we infer a $\tmin$ for CC of $\tmin < \minage$
Myr, a $\tmax$ of $\tmax = \maxage$ Myr, and power-law slope in
between of $\beta = \betavalue$.  Assuming single-star evolution, this
age distribution corresponds to a progenitor mass distribution with a
$\minmass$ of $\minmass$~=~$\minmassvalue$~$\msun$, a slope of
$\alpha$~=~$\alphavalue$, and a $\maxmass$ of 
$\maxmass < \maxmassvalue$~$\msun$ (see Figure~\ref{fig:M31M33bayesmasses}).  

The $\minmass$ is consistent with the estimates from direct-imaging
surveys.  Since there are far more local SNRs than local SNe, the
precision is much higher. Within our detection limits, we find no
evidence for an upper mass. However, we do infer a progenitor mass
distribution that is somewhat steeper than the Salpeter initial
mass function. Either SNR catalogs are significantly biased against finding SNRs in the youngest SF regions, or the most massive stars are not exploding as often as the lowest masses.


Another major assumption of this work is assuming
single-star evolution in transforming the progenitor age distribution
to a mass distribution \citep{marigo2017}. Under different assumptions, the parameters may systematically shift.  For example, \citet{zapartas2017} suggest that mass transfer and mergers in binary evolution could increase $\tmax$ for CCSNe. This issue has also been explored by \citet{maund2017,maund2018} in which they argue that even though their derived ages from resolved stellar populations are consistent with those derived from measurements of H$\alpha$ for nearby H~II regions using single-star stellar population synthesis models, it is possible that they could potentially be in disagreement with binary population synthesis models. Given the single-star assumption of this manuscript, these binary effects would lead to a lower $\minmass$ inference. The Bayesian framework developed here easily allows for other models, including the delayed distributions caused by binary evolution. In fact, the Bayesian evidence will provide a means to estimate whether single-star or binary models best represent the progenitor age distribution.

\section*{Acknowledgements}
Based on observations made with the NASA/ESA Hubble Space Telescope,
obtained [from the Data Archive] at the Space Telescope Science
Institute, which is operated by the Association of Universities for
Research in Astronomy, Inc., under NASA contract NAS 5-26555. Support for programs
\# HST-AR-13882, \# HST-AR-15042, and \# HST-GO-14786 was provided by NASA through a grant from
  the Space Telescope Science Institute, which is operated by the
  Association of Universities for Research in Astronomy, Inc., under
  NASA contract NAS 5-26555.

 \software{emcee \citep{foreman-mackey2013}, DOLPHOT \citep{dolphin2002,dolphin2012,dolphin2013}}



\bibliographystyle{apj}

\end{document}